\documentclass[twocolumn,showpacs,amsmath,amssymb,prc,superscriptaddress,floatfix,aps]{revtex4}
\usepackage{amssymb}
\usepackage{amsmath}
\usepackage{graphicx}
\usepackage{hyperref}
\usepackage{booktabs}

\usepackage{color}
\begin{document}
\title{Total and differential cross sections of the $\boldsymbol{dp\to {}^3}\textrm{He}\,\boldsymbol{\eta}$ reaction at excess energies between 1 and 15~MeV}
%

\author{C.~Fritzsch}\email[E-mail: ]{c.fritzsch@uni-muenster.de}
\affiliation{Institut f\"ur Kernphysik, Westf\"alische Wilhelms-Universit\"at M\"unster, Wilhelm-Klemm-Str. 9, 48149 M\"unster, Germany}
\author{S.~Barsov}
\affiliation{High Energy Physics Department, St. Petersburg Nuclear Physics Institute, RU-188350 Gatchina, Russia}
\author{I.~Burmeister}
\affiliation{Institut f\"ur Kernphysik, Westf\"alische Wilhelms-Universit\"at M\"unster, Wilhelm-Klemm-Str. 9, 48149 M\"unster, Germany}
\author{S.~Dymov}
\affiliation{Institut f\"ur Kernphysik, Forschungszentrum J\"ulich, D-52425 J\"ulich, Germany}
\affiliation{Laboratory of Nuclear Problems, JINR, RU-141980 Dubna, Russia}
\author{R.~Gebel}
\affiliation{Institut f\"ur Kernphysik, Forschungszentrum J\"ulich, D-52425 J\"ulich, Germany}
\author{M.~Hartmann}
\affiliation{Institut f\"ur Kernphysik, Forschungszentrum J\"ulich, D-52425 J\"ulich, Germany}
\author{A.~Kacharava}
\affiliation{Institut f\"ur Kernphysik, Forschungszentrum J\"ulich, D-52425 J\"ulich, Germany}
\author{A.~Khoukaz}\email[E-mail: ]{khoukaz@uni-muenster.de}
\affiliation{Institut f\"ur Kernphysik, Westf\"alische Wilhelms-Universit\"at M\"unster, Wilhelm-Klemm-Str. 9, 48149 M\"unster, Germany}
\author{V.~Komarov}
\affiliation{Laboratory of Nuclear Problems, JINR, RU-141980 Dubna, Russia}
\author{P.~Kulessa}
\affiliation{H.~Niewodniczanski Institute of Nuclear Physics PAN, PL-31342 Cracow, Poland}
\author{A.~Kulikov}
\affiliation{Laboratory of Nuclear Problems, JINR, RU-141980 Dubna, Russia}
\author{A.~Lehrach}
\affiliation{Institut f\"ur Kernphysik, Forschungszentrum J\"ulich, D-52425 J\"ulich, Germany}
\author{B.~Lorentz}
\affiliation{Institut f\"ur Kernphysik, Forschungszentrum J\"ulich, D-52425 J\"ulich, Germany}
\author{D.~Mchedlishvili}
\affiliation{Institut f\"ur Kernphysik, Forschungszentrum J\"ulich, D-52425 J\"ulich, Germany}
\affiliation{High Energy Physics Institute, Tbilisi State University, GE-0186 Tbilisi, Georgia}
\author{T.~Mersmann}
\affiliation{Institut f\"ur Kernphysik, Westf\"alische Wilhelms-Universit\"at M\"unster, Wilhelm-Klemm-Str. 9, 48149 M\"unster, Germany}
\author{M.~Mielke}
\affiliation{Institut f\"ur Kernphysik, Westf\"alische Wilhelms-Universit\"at M\"unster, Wilhelm-Klemm-Str. 9, 48149 M\"unster, Germany}
\author{S.~Mikirtychiants}
\affiliation{High Energy Physics Department, St. Petersburg Nuclear Physics Institute, RU-188350 Gatchina, Russia}
\author{M.~Nioradze}
\affiliation{High Energy Physics Institute, Tbilisi State University, GE-0186 Tbilisi, Georgia}
\author{H.~Ohm}
\affiliation{Institut f\"ur Kernphysik, Forschungszentrum J\"ulich, D-52425 J\"ulich, Germany}
\author{M.~Papenbrock}
\affiliation{Institut f\"ur Kernphysik, Westf\"alische Wilhelms-Universit\"at M\"unster, Wilhelm-Klemm-Str. 9, 48149 M\"unster, Germany}
\author{D.~Prasuhn}
\affiliation{Institut f\"ur Kernphysik, Forschungszentrum J\"ulich, D-52425 J\"ulich, Germany}
\author{V.~Serdyuk}
\affiliation{Institut f\"ur Kernphysik, Forschungszentrum J\"ulich, D-52425 J\"ulich, Germany}
\author{H.~Str\"oher}
\affiliation{Institut f\"ur Kernphysik, Forschungszentrum J\"ulich, D-52425 J\"ulich, Germany}
\author{A.~T\"aschner}
\affiliation{Institut f\"ur Kernphysik, Westf\"alische Wilhelms-Universit\"at M\"unster, Wilhelm-Klemm-Str. 9, 48149 M\"unster, Germany}
\author{Yu.~Valdau}
\affiliation{Institut f\"ur Kernphysik, Forschungszentrum J\"ulich, D-52425 J\"ulich, Germany}
\affiliation{High Energy Physics Department, St. Petersburg Nuclear Physics Institute, RU-188350 Gatchina, Russia}
\author{C.~Wilkin}
\affiliation{Physics and Astronomy Department, UCL, Gower Street, London WC1E 6BT, United Kingdom}

\collaboration{ANKE Collaboration}
\date{\today}
\begin{abstract}
New high precision total and differential cross sections are reported for the $dp\to {}^3\textrm{He}\,\eta$ reaction close to threshold. The measurements were performed using the magnetic spectrometer ANKE, which is an internal fixed target facility at the COSY cooler synchrotron. The data were taken for deuteron beam momenta between $3.14641~\textrm{GeV}/c$ and $3.20416~\textrm{GeV}/c$, which corresponds to the range in excess energy $Q$ for this reaction between $1.14~\textrm{MeV}$ and $15.01~\textrm{MeV}$. The normalization was established through the measurement in parallel of deuteron-proton elastic scattering and this was checked through the study of the $dp\to {}^3\textrm{He}\,\pi^0$ reaction. The previously indicated possible change of sign of the slope of the differential cross sections near the production threshold, which could be explained by a rapid variation of the $s$- and $p$-wave interference term, is not confirmed by the new data. The energy dependence of the total cross section and the $90^{\circ}$ slope parameter are well explained by describing the final state interaction in terms of a complex Jost function and the results are significant in the discussion of $\eta$-mesic nuclei. In combination with recently published WASA-at-COSY data [P.~Adlarson \emph{et al.}, Phys.\ Lett.\ B \textbf{782}, 297 (2018)], a smooth variation of the slope parameter is achieved up to an excess energy of $80.9~\textrm{MeV}$.
\end{abstract}
\maketitle

%
%
\section{Introduction}

Initial measurements at Saclay of the cross section for the $dp\to {}^3\textrm{He}\,\eta$ reaction near threshold~\cite{Berger:1988} could be most easily understood if there were a pole in the $\eta\,{}^3\textrm{He}$ elastic scattering amplitude~\cite{Wilkin:1993} at low excess energy $Q$, which is the kinetic energy in the $\eta\,{}^3\textrm{He}$ center-of-mass system. There had already been suggestions that the interaction of the $\eta$ meson with nucleons was strongly attractive~\cite{Bhalerao:1985cr} and these led Haider and Liu in 1986~\cite{Haider:1986EtaMesicNuclei} to predict a novel state of nuclear matter, where an $\eta$ meson is bound to a nucleus. Due to uncertainties in the strength of the $\eta$-nucleon interaction, they suggested that $\eta$-nuclei would only be formed for nuclei starting from $^{12}$C. Nevertheless, an anomalous behavior has also been observed in the photoproduction reaction of the same $^3\textrm{He}\,\eta$ final state in $\gamma {}^3\textrm{He} \to {}^3\textrm{He}\,\eta$~\cite{Pfeiffer:2004,Pfeiffer:2005,Pheron:2012}, though these data are only available in 4~MeV bins.

\begin{figure}[!h]
	\centering
	\includegraphics[width=0.5\textwidth]{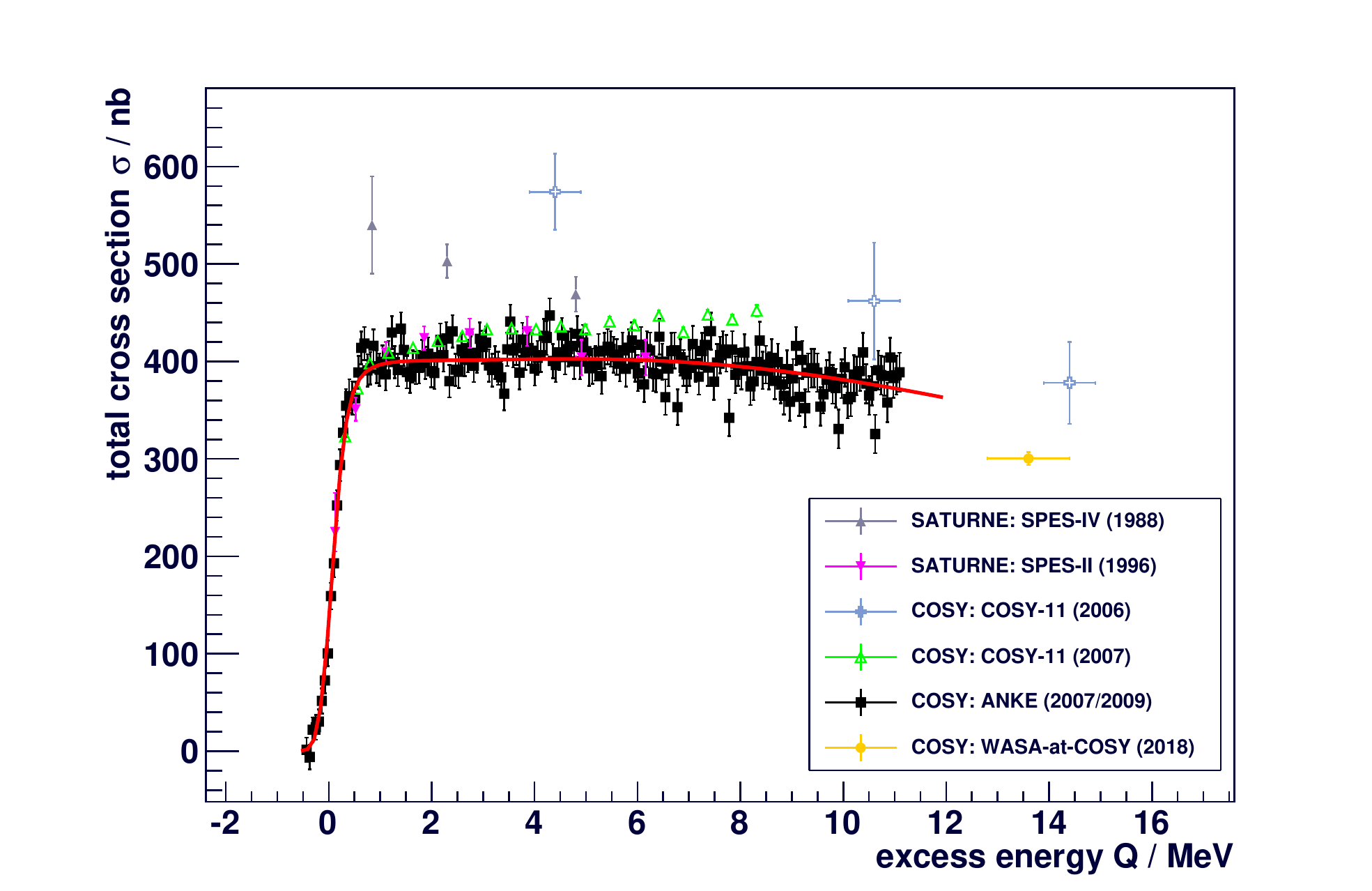}
	\caption{(Color online) Published total cross section data of the $dp \to {}^3\textrm{He}\,\eta$ reaction~\cite{Berger:1988,Mersmann:2007,Smyrski:2007,Adlarson:2018,Mayer:1995,Adam:2007} as a function of excess energy $Q$. The red line is the result of fitting the ANKE data with a two-pole ansatz, as described in Ref.~\cite{Mersmann:2007}. For clarity, no normalization uncertainties are shown. }
	\label{fig:TotalReference}
\end{figure}

The $dp\to {}^3\textrm{He}\,\eta$ results were confirmed in much more refined measurements at ANKE~\cite{Mersmann:2007} and COSY-11~\cite{Smyrski:2007}. It should be noted that the two data sets are completely compatible but the parameters deduced are different because the close-to-threshold ANKE analysis~\cite{Mersmann:2007} required consideration of a non-negligible spread in beam momentum. The near-threshold energy dependence of the total cross section in all the published data is shown in Fig.~\ref{fig:TotalReference}. The rapid rise from threshold is much steeper than that expected from phase space and it is this which is interpreted as being due to a strong $\eta\,{}^3\textrm{He}$ final state interaction (FSI) and the possible formation of a (quasi)-bound state of the $\eta\,{}^3\textrm{He}$-system close to the production threshold~\cite{Wilkin:1993}. For excess energies above about 1~MeV the total cross section seems to reach a plateau at a level of about $\sigma = 400$~nb and this suggests that the FSI pole must lie in the complex plane with $|Q| \lesssim 1$~MeV. To account for the high and low $Q$ behavior, the ANKE data were fitted with an FSI factor that was the product of two poles~\cite{Mersmann:2007} and this showed that the nearby pole was indeed at $|Q| < 1$~MeV with relatively small errors in both the real and imaginary parts of $Q$.

\begin{figure}[!h]
	\centering
	\includegraphics[width=0.5\textwidth]{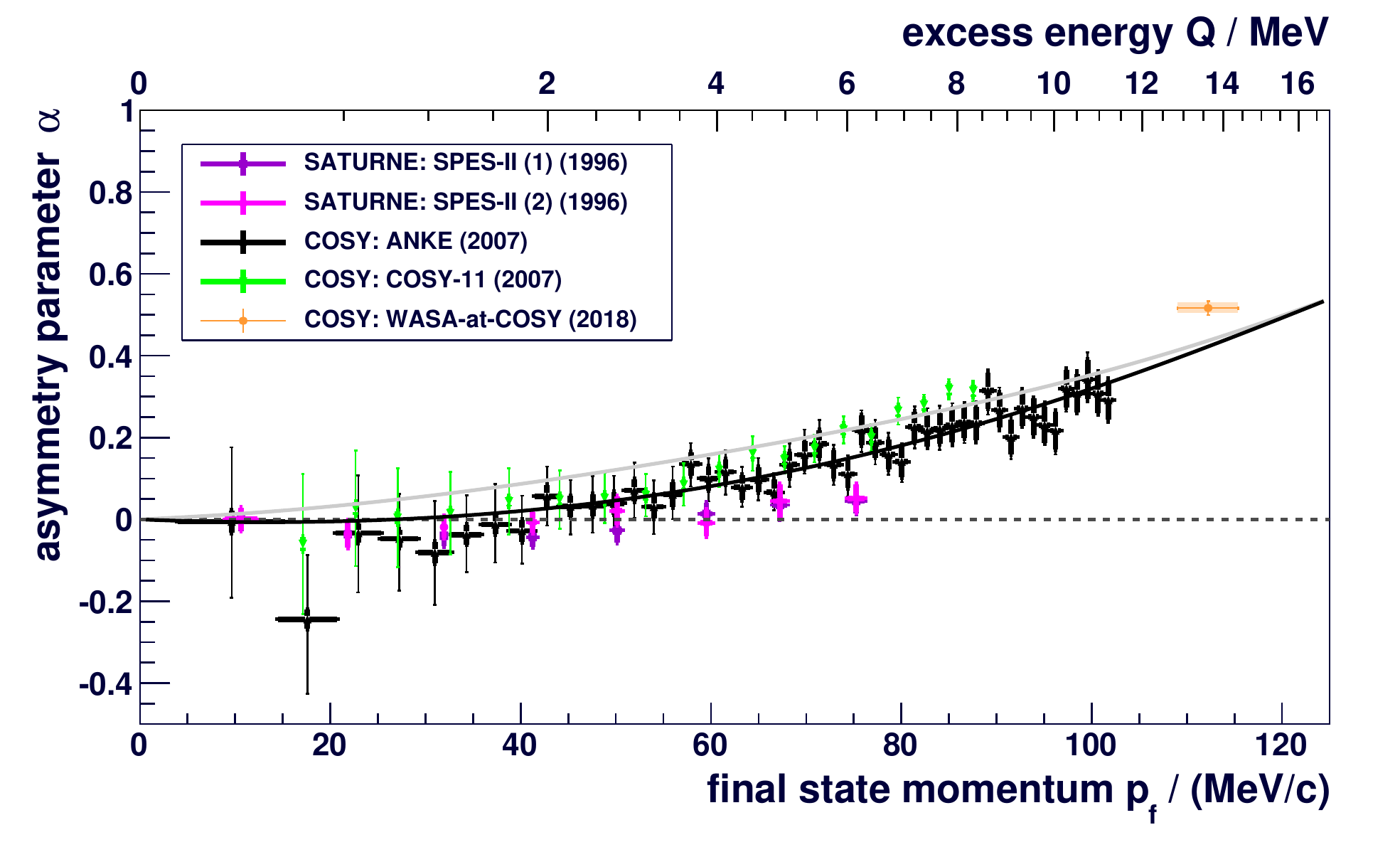}
	\caption{(Color online) Published values of the asymmetry parameter $\alpha$ of Eq.~\eqref{alpha} as a function of the final center-of-mass momentum $p_f$~\cite{Mersmann:2007,Smyrski:2007,Adlarson:2018,Mayer:1995,Adam:2007}. The black solid line is the result of fitting the ANKE data including the momentum dependence of the $s$- and $p$-wave interference term. The light gray line is obtained by assuming that the relative phase of the $s$- and $p$-wave amplitudes is independent of momentum. A detailed description of the fits is given in Ref.~\cite{Wilkin:2007}.}
	\label{fig:AsymReference}
\end{figure}

In the comparison of data obtained at different facilities, the biggest uncertainty is in the absolute normalization of the cross sections. This is avoided when one looks at the logarithmic slope at $90^{\circ}$, defined as
\begin{equation}
	\label{alpha}
	\alpha = \frac{d}{dz}\left\{\ln{\bigg( \frac{d\sigma}{d\Omega} \bigg)}\right\}   \bigg|_{z=0}
           = \left.\frac{d}{dz}{\bigg( \frac{d\sigma}{d\Omega} \bigg)}\right/\bigg( \frac{d\sigma}{d\Omega} \bigg)\bigg|_{z=0},
\end{equation}
where $z=\cos\vartheta$, with $\vartheta$ being the angle between the momentum of the $\eta$ meson and that of the initial proton (or between the deuteron and $^3$He) in the center-of-mass frame. The published values of the asymmetry parameter $\alpha$ are shown in Fig.~\ref{fig:AsymReference} in terms of the $\eta\,{}^3\textrm{He}$ center-of-mass momentum $p_f$ where, non-relativistically, $Q=p_f^{\,\,2}/2m_{\rm red}$, with $m_{\rm red}$ being the $\eta\,{}^3\textrm{He}$ reduced mass.

Non-zero values of $\alpha$ arise from interference between odd and even partial waves which, at low energies, means principally $s$- and $p$-waves. Since near threshold the $p$-wave amplitude increases like $p_f$, one might then expect that $\alpha$ should show a similar behavior. The experimental data presented in Fig.~\ref{fig:AsymReference} do show a linear dependence but only for $p_f \gtrsim 40$~MeV/$c$ and the values of $\alpha$ might even go negative in the region below 40~MeV/$c$. It was pointed out~\cite{Wilkin:2007} that there would be deviations from this expected linear behavior of $\alpha$  due to the momentum dependence of the $s$-wave amplitude and the position of the pole in the complex $Q$ plane found in the ANKE fit~\cite{Mersmann:2007} would suppress the low $p_f$ values of $\alpha$, as shown in Fig.~\ref{fig:AsymReference}. Although this kind of production data will never distinguish between bound and antibound systems, the real and imaginary parts of the pole position are of great importance in the study of the development of $\eta$ mesic-nuclei.

In view of the importance of the $\eta\,{}^3\textrm{He}$ system for the understanding of $\eta$ mesic-nuclei, it is helpful to repeat the measurements of the $dp \to {}^3\textrm{He}\,\eta$ cross section at low energies with high statistics and better determination of the kinematics. The opportunity arose to use the data acquired for the measurement of the mass of the $\eta$ meson. Values of the deuteron beam momenta were first obtained by taking polarized beams and studying the position of an artificially induced depolarizing resonance~\cite{Goslawski:2010}. The $\eta$ mass was then determined to very high accuracy purely from kinematics, using the locations of the $^3$He hits on the ANKE focal plane~\cite{Goslawski:2012}. The energy dependence of the cross section was not used in this analysis, though the statistical precision benefitted from the high cross section near threshold that is apparent in Fig.~\ref{fig:TotalReference}. Since the experimental methodology is well explained in these papers, the description of the experiment in Sec.~\ref{experiment} can be quite brief.

Though the experimental considerations are identical to those in the $\eta$-mass experiment~\cite{Goslawski:2010,Goslawski:2012}, once the $dp \to {}^3\textrm{He}\,\eta$ reaction has been identified, the analysis of the data described in Sec~\ref{Analysis} is very different because here we are interested in studying the count rates. In order to convert count rates to cross sections it is necessary to establish the normalization, i.e., the luminosity. For internal experiments in a storage ring such as COSY it is standard to compare the corrected count rates with those of a reaction whose normalization is known. The reaction chosen for this purpose is deuteron-proton elastic scattering, which has been well studied in the energy region required for this experiment~\cite{Fritzsch:2018,Fritzsch:2019}. As discussed in Sec.~\ref{Normalization}, the reaction is easily identified, with a high cross section that is weakly energy dependent. An independent check on the normalization is provided through the measurement of the $dp\to {}^3\textrm{He}\,\pi^0$ reaction, which is discussed in Sec.~\ref{NormCrossCheck}. Though the limitations on the published database make this method less precise than deuteron-proton elastic scattering, it does show clearly that we have not generated a false energy behavior through the luminosity assumptions.

Our results for the differential and total cross sections for $\eta$ production are presented in Sec.~\ref{Results}. The striking rise of the total cross section over 1~MeV in excess energy indicates that there must be a final state interaction pole for $|Q|\lesssim 1$~MeV, though its position in the complex $Q$ plane is much more model dependent. Though there are clear differences between the angular distributions measured at our highest energy and the lowest energies studied in the WASA-at-COSY experiment~\cite{Adlarson:2018}, the slope parameter varies smoothly over a wide range of final state momentum. Nevertheless, the behavior of the slope near threshold is markedly different to that found in some of the earlier experiments which had been explained as being due a strong energy dependence of the $s$-wave amplitude~\cite{Wilkin:2007}. In particular there is no sign in our data of the slope changing sign at low energies. The earlier ANKE data on the total cross section and slope parameter~\cite{Mersmann:2007} were modelled with a final state interaction that was the product of two poles~\cite{Wilkin:2007} but the modelling would be on much firmer ground if it were taken as the product of a zero and a pole, as is done in Sec.~\ref{FSI}. Though the magnitude of the pole position $|Q|$ is little changed from the 1~MeV found earlier~\cite{Mersmann:2007,Wilkin:2007}, the phase is changed and this difference is significant in the context of the $\eta$-mesic nucleus discussion. The final section, Sec.~\ref{Summary}, tries to summarize the results and put them into some kind of context, especially within the discussion of $\eta$-mesic nuclei more broadly.
%
%
\section{Experiment}
\label{experiment}

\begin{figure}[h]
	\centering
	\includegraphics[width=0.95\columnwidth]{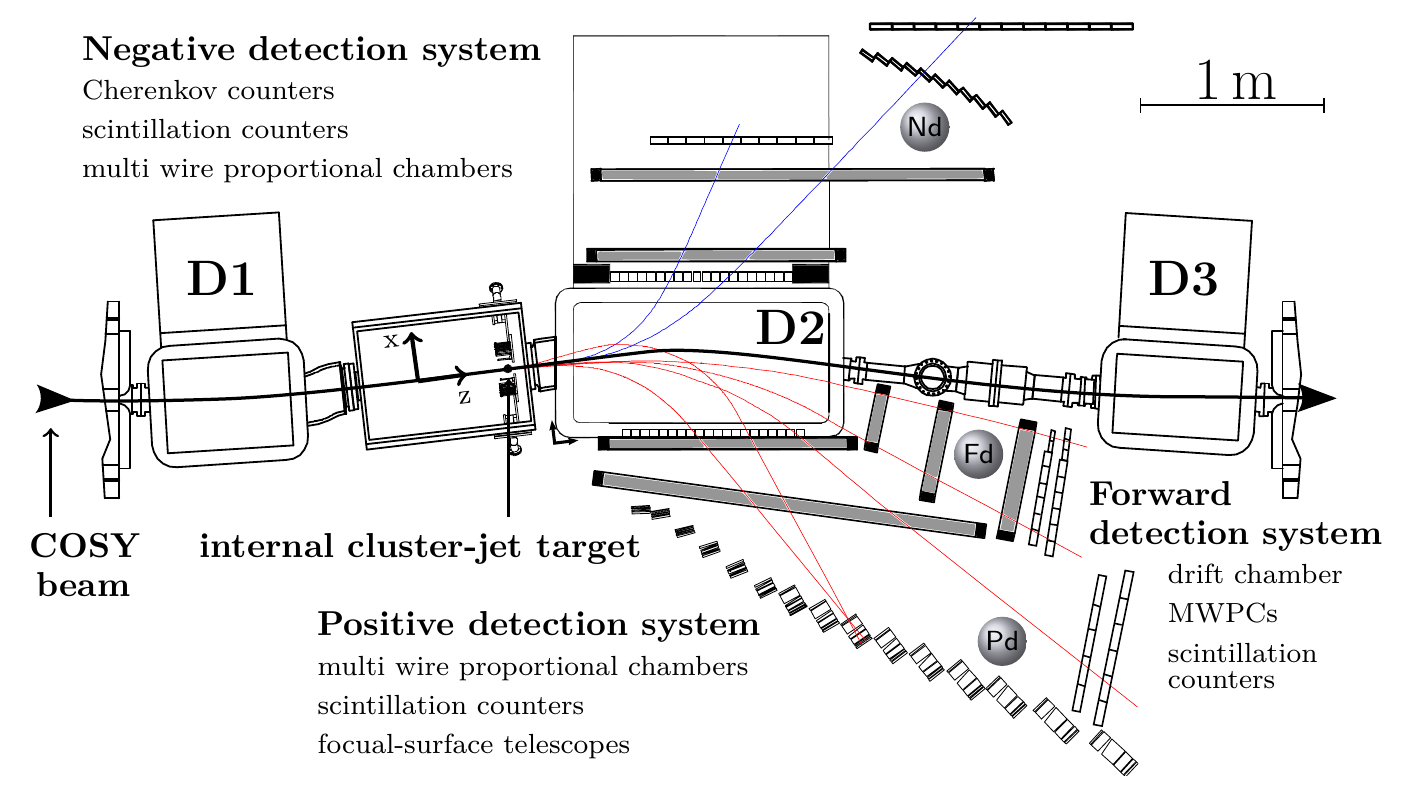}
	\caption{(Color online) Schematic view of the magnetic spectrometer ANKE \cite{Barsov:2001}. The main components are the three dipole magnets D1--D3, the internal cluster-jet target, and the three different detection systems (Fd, Pd and Nd), though only the forward detector was used in this experiment. Typical trajectories of negatively charged particle are shown by blue lines, whereas examples of positively charged ones are sketched as red lines.}
	\label{fig:anke}
\end{figure}

As mentioned in the Introduction, the data presented here were by-products of an experiment to measure the mass of the $\eta$ meson to high accuracy~\cite{Goslawski:2010,Goslawski:2012}. The description of the experiment itself can therefore be relatively concise.
The data were taken using the magnetic spectrometer ANKE~\cite{Barsov:2001}, which is an internal fixed target facility situated inside the cooler synchrotron storage ring (COSY) of the For\-schungs\-zen\-t\-rum J\"ulich. A schematic overview of ANKE is shown in Fig.~\ref{fig:anke}.

The measurements were performed using an unpolarized deuteron beam incident on a hydrogen cluster-jet target~\cite{Khoukaz:1999}. The experiment was carried out in the so-called SuperCycle (SC) mode of COSY, where the beam is alternated between up to eight different momentum settings, called FlatTops (FT). The advantage of the supercycle mode is that systematic effects between different beam momentum settings are minimized. To cover the range of deuteron momenta $p_d$ between 3120.17 and 3184.87~MeV/$c$ three supercycles were used. The accuracy of $\Delta p_{\rm d}/p_{\rm d} < 6 \times 10^{-5}$ was determined by using the spin-depolarization technique~\cite{Goslawski:2010}. In total there were 15 different excess energies $Q$ with respect to the $dp \to {}^3\textrm{He}\,\eta$ threshold plus two measurements below threshold to control the background. The values of the excess energies for each of the different beam momentum settings are given in Table~\ref{tab:BeamMomenta}.

Following the interaction of the deuterons with the cluster-jet target~\cite{Khoukaz:1999}, the ejectiles produced are separated by the dipole magnet D2 according to their rigidity. A special feature of the magnetic spectrometer ANKE is the movable D2 analyzing magnet, which can be shifted transversely to the COSY beam direction to optimize the geometrical acceptance of each reaction being investigated. Due to fixed-target kinematics, the positively charged heavy ${}^3\textrm{He}$ nuclei are boosted in the forward direction into the acceptance of the forward detector (Fd-system). This consists of one multi-wire drift and two multi-wire proportional chambers (used for track reconstruction) and three layers of plastic scintillators (used for energy loss and time-of-flight measurements). In general, the third layer is part of the so called side wall of the positive detection system but in order to improve the ${}^3\textrm{He}$ nuclei selection of this measurement it was moved behind the two scintillator layers of the forward detection system.
Since the $\eta$-meson has no charge, having identified the $^3$He the $\eta$ production is inferred from the missing mass in the reaction.

\begin{table}[h!]
	\centering
	\caption{Excess energy $Q$ in MeV with respect to the reaction $dp \to {}^3\textrm{He}\,\eta$ for each supercycle and flattop. The statistical uncertainty of the excess energy is $\Delta Q_{\rm{stat}} = 0.01~\textrm{MeV}$ and the systematic uncertainty is $\Delta Q_{\rm{sys}} = 0.03~\textrm{MeV}$ for each flattop. }
	\scriptsize
	\begin{tabular}{p{0.5cm}|p{0.65cm}p{0.65cm}p{0.65cm}p{0.65cm}p{0.65cm}p{0.65cm}p{0.65cm}}		
		\toprule
		& $Q_{\mathrm{FT1}}$  & $Q_{\mathrm{FT2}}$ & $Q_{\mathrm{FT3}}$  & $Q_{\mathrm{FT4}}$  & $Q_{\mathrm{FT5}}$  & $Q_{\mathrm{FT6}}$  & $Q_{\mathrm{FT7}}$  \\
		\specialrule{2pt}{4pt}{6pt}
		SC1 & --5.15 & 1.14 & 1.63 & 2.59 & 4.09 & 6.33 & ~~8.60 \\
		SC2 & --5.15 & 1.36 & 2.10 & 3.08 & 5.07 & 7.32 & 10.37  \\
		SC3 & &  &  & 3.79 & 4.55 &  & 15.01 \\
		\bottomrule
	\end{tabular}
	\label{tab:BeamMomenta}
\end{table}

%
%
\section{Data Analysis}
\label{Analysis}

Since the $\eta$ meson is neutral and it is not possible to detect its decay products in ANKE, the isolation of the ${}^3\textrm{He}\,\eta$ final state relies on a good measurement of the $^3$He in the Fd detector and the subsequent identification of the meson using the missing-mass technique.

\begin{figure}[t]
	\centering
	\includegraphics[width=1\linewidth]{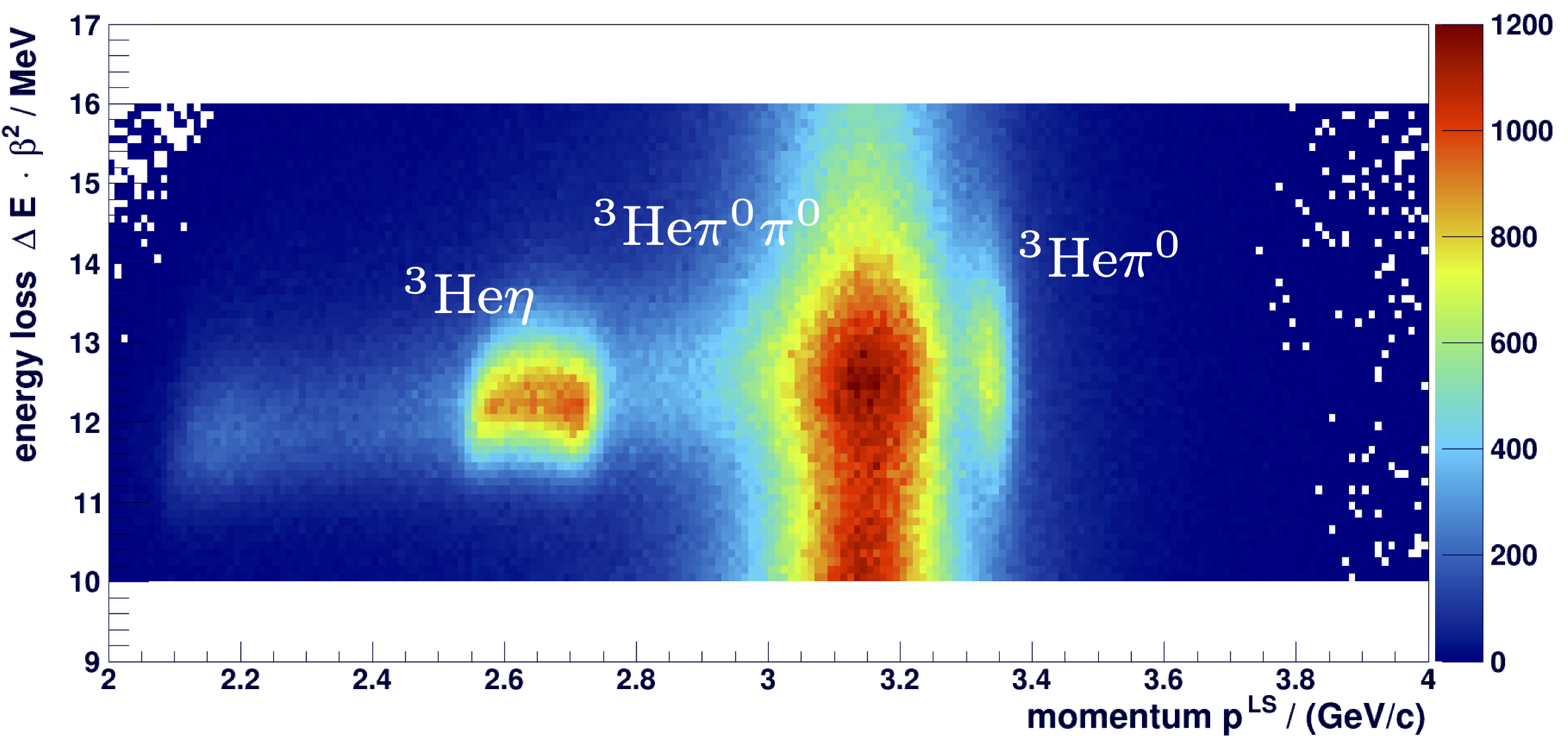}
	\includegraphics[width=1\linewidth]{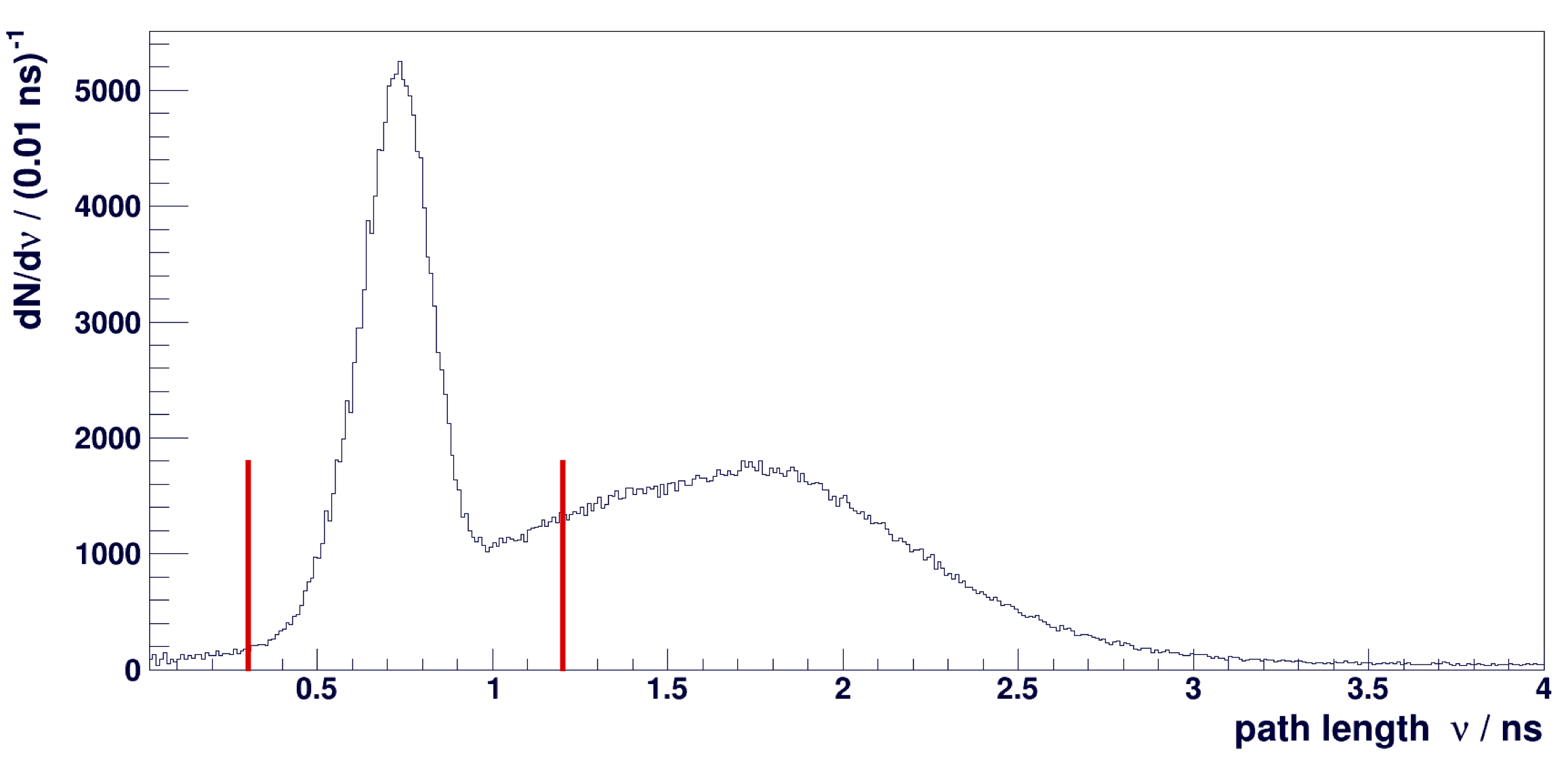}
	\caption{(Color online) {\bfseries Top:} Energy-loss distribution $\beta^2\Delta E$ of the ${}^3\textrm{He}$ ions as a function of the laboratory momentum $p^{LS}$ for reconstructed data at an excess energy of $Q=5.07~\textrm{MeV}$. The energy losses are the sums over all scintillator detectors of the first layer, corrected for their different thicknesses. The horizontal cuts at $\beta^2\Delta E = 10~\textrm{MeV}$, and $16~\textrm{MeV}$ chosen for the analysis do not eliminate any good $\eta\,{}^3\textrm{He}$ events. {\bfseries Bottom:}  Distribution of track length $\nu$ at an excess energy of $Q=5.07~\textrm{MeV}$ determined using the third scintillator layer for the stop signal. The red vertical lines represent the $\pm 4\sigma$ limits around the Gaussian-like peak. }
	\label{fig:EnergyLoss}
\end{figure}

The data presented here were taken using a hardware trigger requiring one coincident hit in each plastic scintillator layer with a significantly higher energy deposit than that expected for deuterons or protons, which dominate the background. By plotting the energy $\Delta E$ of the ${}^3\textrm{He}$ nuclei deposited within the detector material as a function of the reconstructed laboratory momentum $p^{LS}$, a specific energy-loss band for each particle species can be observed. Fig.~\ref{fig:EnergyLoss} (top) shows a typical example of such an energy-loss distribution for data at an excess energy of $Q=5.07~\textrm{MeV}$. The cuts on the energy loss were chosen to be far away from the reaction signal in order to avoid influencing the signal region.

In order to further reduce the amount of competing reactions, a ${}^3\textrm{He}$ track length cut was also applied using the time information from the plastic scintillator layers. The track length $\nu$ was determined by multiplying the time-of-flight by the relativistic velocity $\beta$ of the ${}^3\textrm{He}$ nuclei. Since the first two scintillator layers were placed only around $7~\textrm{cm}$ apart, an accurate determination of the path length between them is not possible. Instead, the average of the times of the first two layers was used as a start signal and that from the third layer as the stop signal. As a typical example, Fig.~\ref{fig:EnergyLoss} (bottom) shows the track length distribution at an excess energy of $Q=5.07~\textrm{MeV}$.
There is a Gaussian-like peak at $\nu \approx 0.7~\textrm{ns}$ and the red vertical lines represent the $\pm 4\sigma$ limits that were used as track length cut values.

\begin{figure}[pt]
	\centering
	\includegraphics[width=1.\linewidth]{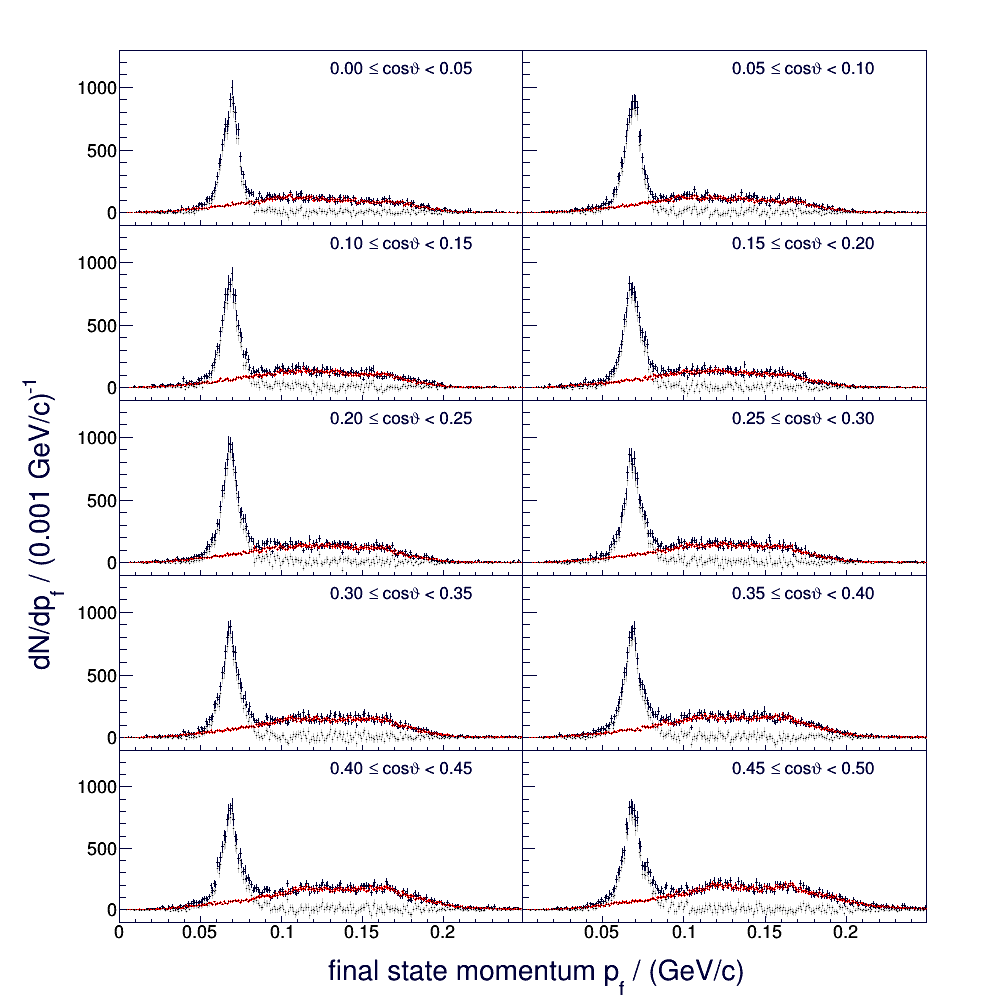}
	\includegraphics[width=1.\linewidth]{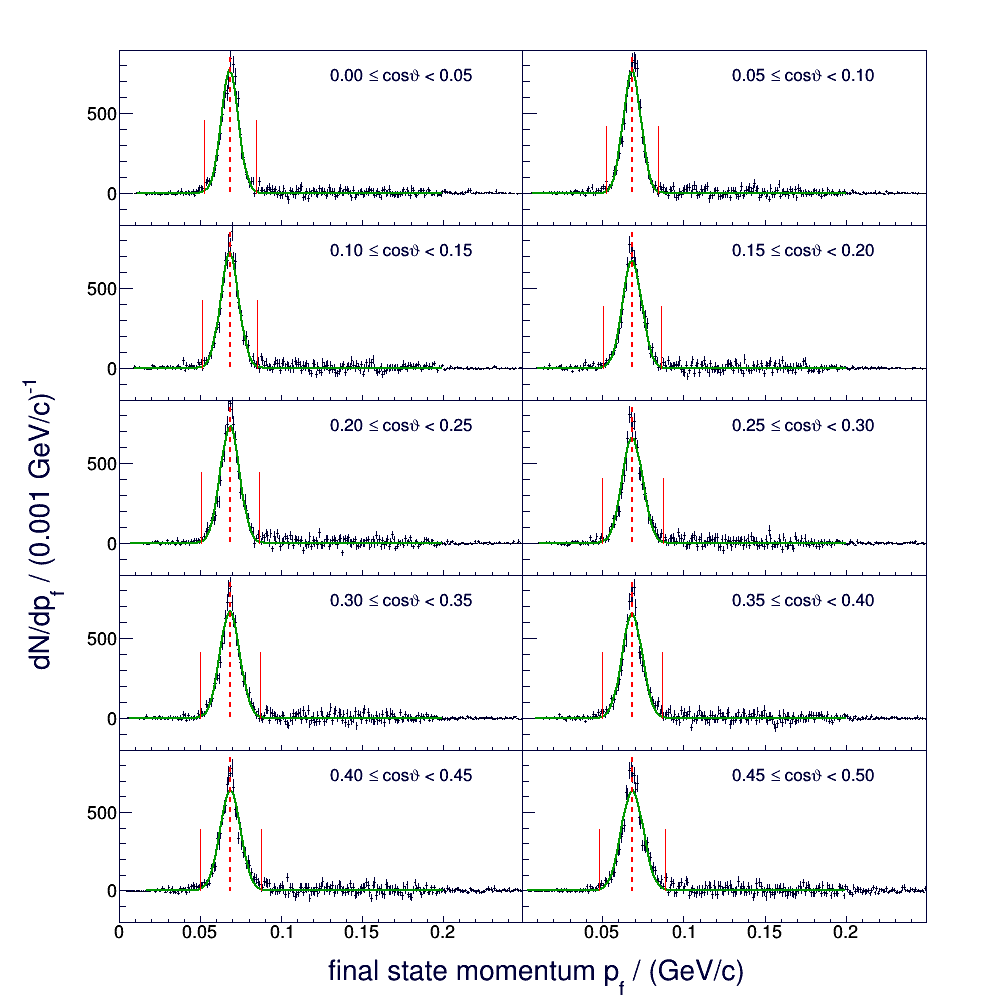}
	\caption{{(Color online) \bfseries Top:} Final state momentum spectra for exemplary $\cos \vartheta$ bins between $0.00 \leq\cos \vartheta < 0.50$ at an excess energy of $Q=5.07~\textrm{MeV}$. Here $\vartheta$ is the center-of-mass angle between the proton and $\eta$ meson, i.e., between the deuteron and $^3$He. The black distribution represents the reaction signal plus the background spectrum, the red distribution the scaled background distribution using the sub-threshold data and the light gray distribution the background-subtracted signal. {\bfseries Bottom:} Background-subtracted final state momentum spectra in the same $\cos \vartheta$ intervals. The green line is the result of fitting a Gaussian and the full red lines the $\pm 3\sigma$ range of the Gaussian fit. The red dashed line show the position of the peak expected when using the accepted value of the mass of the $\eta$ meson~\cite{Tanabashi:1898}.}
	\label{fig:BackgroundSignal}
\end{figure}

To eliminate the remaining contributions from other reactions, a model-independent background description was applied. For this purpose, sub-threshold data at an excess energy of $Q=-5.15~\textrm{MeV}$ were collected during the first two supercycles. These data were analyzed on an event-by-event basis as if they had been taken above the ${}^3\textrm{He}\,\eta$ threshold at the specific beam momentum settings of the beam time. This leads to shifts of the kinematic limits in the missing mass or final state momentum distributions for each of the flattops. After correcting for the different luminosities measured at the various beam momentum settings, the scaled background spectrum was subtracted from the above-threshold data. This procedure was carried out for 40 $\cos \vartheta$ bins with a bin width of $\Delta\cos \vartheta = 0.05$ to provide an accurate investigation of the angular dependence (cf.\ Fig.~\ref{fig:BackgroundSignal}, top). The signal yields for each $\cos \vartheta$ bin was determined by summing over the $\pm 3\sigma$ range of a Gaussian fit to the background-subtracted data (cf.\ Fig.~\ref{fig:BackgroundSignal}, bottom).

The geometrical acceptance of the detector was also determined from Monte Carlo simulations. Here, the same cut conditions and counting methods were applied as in the experiment in order to estimate the acceptance-corrected yields for each flattop and $\cos \vartheta$ bin. In general, the geometrical acceptance factor is $90\%$ or higher, except for the highest energy. The 15~MeV data, which were taken well above the ${}^3\textrm{He}\,\eta$ production threshold and close to the limit of full geometrical acceptance of ANKE, have acceptance factors that lie between 50\% and 80\%, depending upon the polar angle of the production. This geometrical acceptance factor must be estimated in an iterative way since detector resolution can cause migration effects, which means that some events will be reconstructed in a different $\cos \vartheta$ bin to the one in which they were generated. In order to correct for this, the distributions of the acceptance corrected number of events were fitted with a polynomial, which served as the input for new Monte Carlo simulations in the next iteration. This procedure was repeated until the distribution of the geometrical acceptance factors converged.
%
%
\section{Normalization through $\boldsymbol{dp}$ elastic scattering}
\label{Normalization}

The normalization of the $dp\to {}^3\textrm{He}\,\eta$ reaction was assured by comparing the corrected count rates with those of deuteron-proton elastic scattering, which was measured in parallel. The advantage of using this reaction is the wide available data base with high differential cross sections $d\sigma/dt$, on the order of $10^4~\mu\textrm{b}/(\textrm{GeV}/c)^2$ over the ANKE acceptance range between $0.08\;(\textrm{GeV}/c)^2 \leq |t| < 0.26\;(\textrm{GeV}/c)^2$. This ensures an excellent signal-to-background ratio.  Here $t$ is the square of the four-momentum transfer and it is important to note that $d\sigma/dt$ has a weak energy dependence over the energy interval required for this experiment~\cite{Fritzsch:2018,Fritzsch:2019}. The small variations, which are not more than $3\%$ at the highest available momentum transfers,  might be due to the uncertainty in the input $NN$ amplitudes that were used for the refined Glauber calculations.

The identification of elastic scattering was accomplished by detecting the fast deuterons in the Fd-system. For this purpose, a second hardware online trigger was applied to handle the enormous amount of data. In contrast to the identification of the ${}^3\textrm{He}$ nuclei, this trigger required at least one hit in each of the first two scintillator layers with low energy deposit, since the deuterons carry half of the charge of ${}^3\textrm{He}$ nuclei. In addition, a pre-scaling factor of 1024 was applied in order to reduce the dead time of the data acquisition system. Due to the low momentum transfer to the target proton, deuterons from elastic scattering have momenta close to that of the beam. A simple cut on the ratio $R$ between the reconstructed deuteron momentum and the nominal beam momentum, $R > 0.913$, removes the vast majority of the background and allows one to investigate an almost background-free elastic scattering signal.

As in the case of the $dp\to {}^3\textrm{He}\,\eta$ reaction, the numbers of elastic scattering events were determined by fitting a Gaussian to the missing-mass spectrum and summing all events within $\pm3\sigma$ of the peak. The determination of the acceptance factors was also made in the same way as described earlier, using Monte Carlo simulations. For each beam momentum setting this procedure was done for 18 momentum transfer bins with a bin width of $\Delta|t| = 0.01\;(\textrm{GeV}/c)^2$. This showed that the acceptance factor drops from 15\% to 7\% with increasing momentum transfer.

The published $dp$ elastic scattering differential cross sections $d\sigma/d t$~\cite{Dalkhazhav:1969,Winkelmann:1980,Irom:1984,Velichko:1988,Guelmez:1991} were fit to the function
\begin{equation}
	f(|t|) = \exp(a + b|t| + c|t|^2)~\mu\textrm{b}/(\textrm{GeV}/c)^2
\end{equation}
in the momentum transfer interval $0.05\;(\textrm{GeV}/c)^2 \leq -t < 0.4\;(\textrm{GeV}/c)^2$. This led to parameters $a = 12.45$, $b = -27.24\;(\textrm{GeV}/c)^{-2}$ and $c = 26.31\;(\textrm{GeV}/c)^{-4}$. This function was integrated over each momentum transfer bin so that the luminosity was determined independently for 18 momentum transfer bins for each beam momentum setting. The results presented in Table~\ref{tab:lumresults} are the weighted mean values of the luminosity for each flattop above the ${}^3\textrm{He}\,\eta$ threshold.
\begin{table}[t]
	\centering
	\caption{Determined luminosities with the statistical uncertainties of the last digits in brackets. The indices $+$ and $-$ on the systematic uncertainties refer to the cases of  higher and lower luminosities, respectively. The first two rows are the results obtained in the below-threshold measurements in supercycles 1 and 2, respectively.}
	\label{tab:lumresults}
	\begin{tabular}{c c c c}
		\toprule
		$Q$   & $L_{\rm{int}}$  & $\Delta L_{\rm{int,sys}}^+$  & $\Delta L_{\rm{int,sys}}^-$  \\
		(MeV) &  ($\textrm{nb}^{-1}$) &  ($\textrm{nb}^{-1}$) &  ($\textrm{nb}^{-1}$) \\
		\specialrule{2pt}{4pt}{6pt}
		--5.15(1) & 2215(24)  & 94 & 94 \\
		--5.15(1) & 2282(24)  & 97 & 104\phantom{1} \\
		\phantom{1}1.14(1) & 1148(13)  & 50 & 49 \\
		\phantom{1}1.36(1) & 1175(16)  & 50  & 56 \\
		\phantom{1}1.63(1) & 1193(13)  & 53  & 54 \\
		\phantom{1}2.10(1) & 1164(12)  & 46 & 78 \\
		\phantom{1}2.59(1) &  1152(14) & 46 & 50\\
		\phantom{1}3.08(1) & 1160(13)  & 49  & 52\\
		\phantom{1}3.79(1) & 1194(14)  & 47 & 54 \\
		\phantom{1}4.09(1) &  1166(13) & 48 & 51 \\
		\phantom{1}4.55(1) & 1209(14)  & 51 & 50 \\
		\phantom{1}5.07(1) & 1191(13)  & 50 & 52\\
		\phantom{1}6.33(1) &  1137(14) & 48 & 50 \\
		\phantom{1}7.32(1) & 1165(14) & 51 & 46 \\
		\phantom{1}8.60(1) & \phantom{1}992(11)  & 42 &  48\\
		10.37(1) & 1054(12)  & 42& 49 \\
		15.01(1) &  \phantom{1}984(11) & 51 &  45 \\
		\bottomrule
	\end{tabular}
\end{table}
With this method a systematic precision of $\Delta L_{\rm{sys}} \approx 6\%$ and statistical precision $\Delta L_{\rm{stat}} \approx 1\%$ were achieved, which are improvements by at least a factor of two compared to the previous measurements at ANKE~\cite{Mersmann:2007,Rausmann:2009}. The systematic uncertainties arise mainly from the absolute normalization of the $dp$ elastic scattering reference data and from possible errors in the setting of the nominal beam deflection angle of ANKE.
%
 %
 \section{Normalization through the $\boldsymbol{dp \rightarrow {}^3\textrm{He}\,\pi^0}$ reaction}
 \label{NormCrossCheck}

 An independent check on the luminosity is provided through the measurement in parallel of the $dp\to{}^3\textrm{He}\,\pi^0$ reaction, whose analysis is described in detail in Ref.~\cite{Fritzsch:2019}. This will also confirm that we have not introduced any spurious energy dependence in the $\eta$ excitation function through some unknown systematic effect in the $dp$ elastic luminosity determination.  A clear advantage of using this reaction is that the identification of the ${}^3\textrm{He}$ nuclei, which differ only in momentum from those shown in Fig.~\ref{fig:EnergyLoss}, is identical to that in the ${}^3\textrm{He}\,\eta$ case, so that all the previously discussed software cuts can be modified and used. After doing this, Fig.~\ref{fig:EnergyLoss} (top) shows a clear island corresponding to $^3\textrm{He}\,\pi^0$ final states.

 In contrast to $\eta$ production, the ${}^3\textrm{He}\,\pi^0$ final state covers the much higher excess energy range between $Q=407.7\;\textrm{MeV}$ and $Q=427.9\;\textrm{MeV}$. The geometrical acceptance of the forward system then restricts the detection of this reaction to near-forward events with $\cos{\vartheta} > 0.86$.

\begin{figure}[t]
 	\centering
 	\includegraphics[width=1.\linewidth]{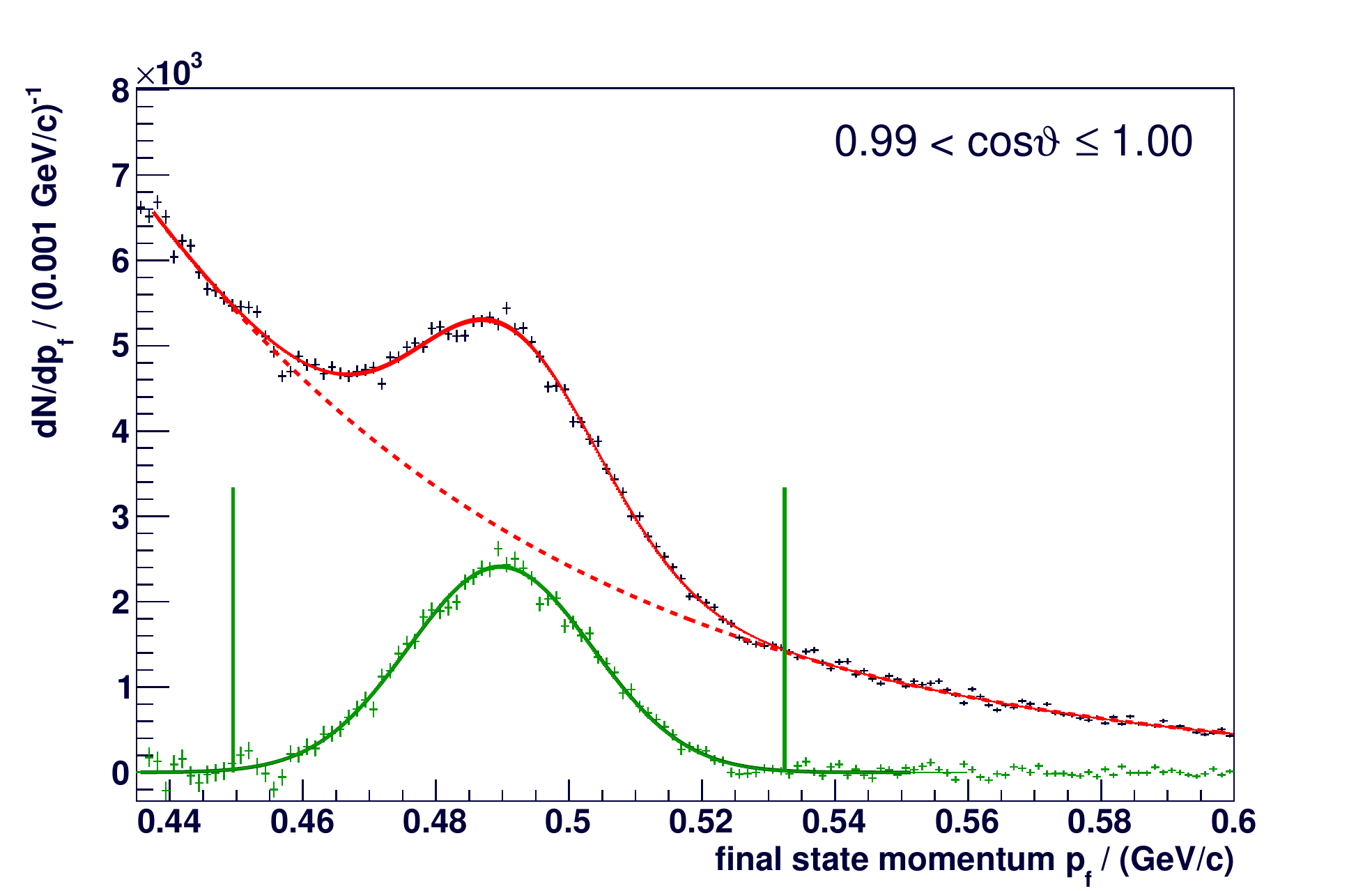}
 	\caption{(Color online) Final state momentum distribution for $0.99 < \cos \vartheta < 1.00$. The red solid line is a fit (background plus Gaussian) to the reconstructed data at $T=1801$~MeV, of which the red dashed line represents just the background. The green solid line is a Gaussian fit to the background-subtracted ${}^3\textrm{He}\,\pi^0$ signal (green histogram) and the green vertical lines represent the $\pm3\sigma$ range assumed for the signal.}
 	\label{fig:BackgroundPion}
 \end{figure}

 One sees in Fig.~\ref{fig:BackgroundPion} a clear peak in the $^3$He momentum distribution corresponding to $\pi^0$ production. This is sitting on a background that arises mainly from
 multi-pion production as well as deuteron breakup reactions. This background was parameterized empirically as
 \begin{equation}
 f(p_f)_{\text{bg}}=  \exp(A\, p_f) \sum_{n=0}^3a_n(p_f)^n.
 \end{equation}
 Taken together with a Gaussian form to represent the $\pi^0$ signal, the data of Fig.~\ref{fig:BackgroundPion} were fit to determine the parameters. Shown separately are the total fit (solid red line), the background (red dashed line), and the background-subtracted spectrum (green distribution). The signal yield was determined by summing over the $\pm3\sigma$ range of a Gaussian fit to the background-subtracted data (solid green line). This procedure was carried out for 14 angular bins from $\cos{\vartheta} = 0.86$ to $\cos{\vartheta} = 1.00$ with uniform widths of $\Delta \cos{\vartheta} = 0.01$. This was then repeated for all 17 beam momentum settings. The geometrical acceptance correction factors were determined in the same way as described in Sec.~\ref{Analysis}. With increasing $\cos{\vartheta}$ the acceptance correction factors rise from 15\% up to 90\% for all energies.

\begin{figure}[t]
 	\centering
 	\includegraphics[width=1.\linewidth]{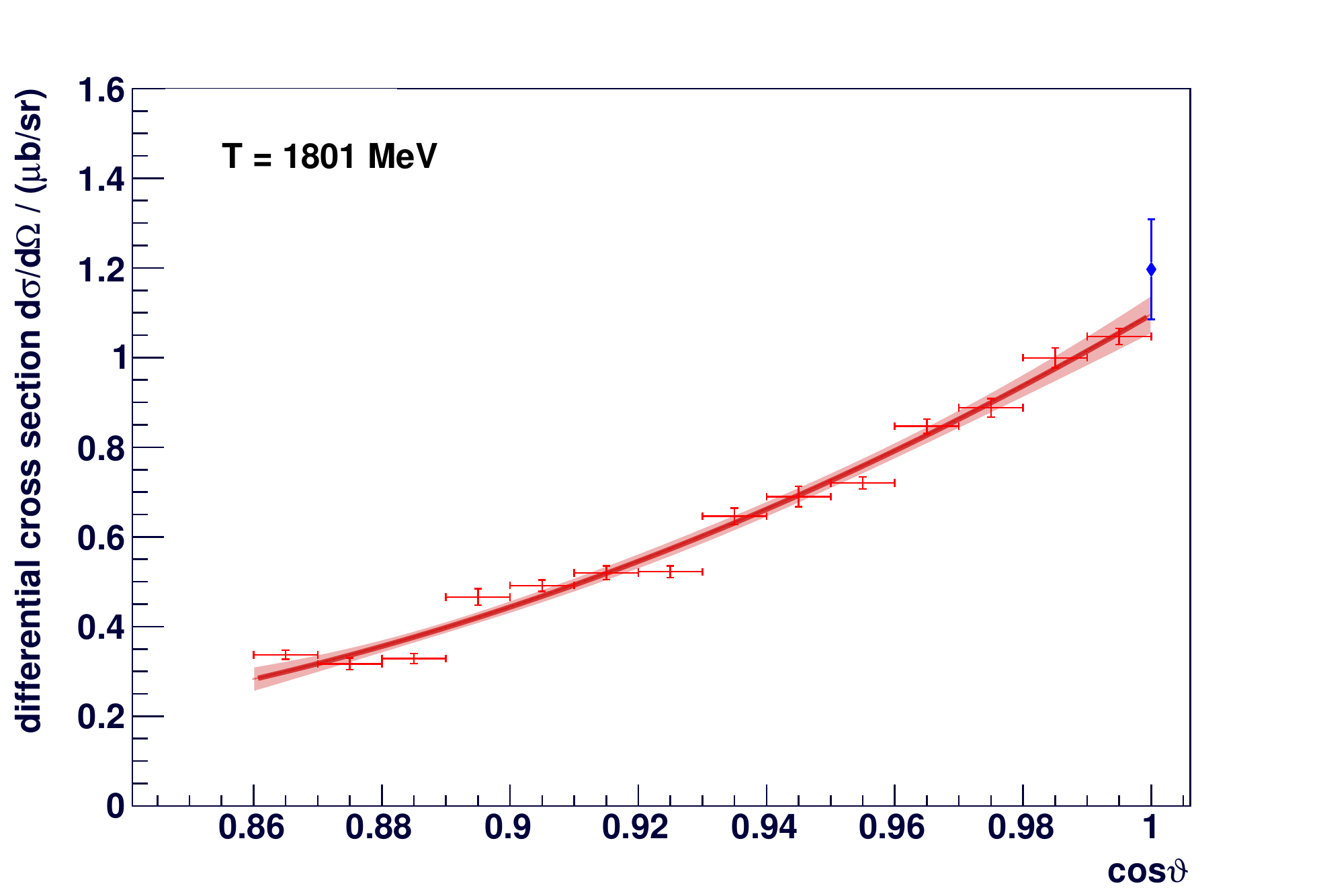}
 	\caption{(Color online) Differential cross section of the $dp \rightarrow {}^3\textrm{He}\,\pi^0$ reaction at small angles obtained at $T=1801$~MeV compared to the interpolation (blue point) of the measurements at zero degrees~\cite{Kerboul:1986}. The uncertainty displayed for the reference point is an amalgam of statistical uncertainty of the interpolation, the $8\%$ normalization uncertainty of the reference data~\cite{Kerboul:1986}, and the systematic uncertainty in the $dp$ elastic luminosities shown in Table~\ref{tab:lumresults}). The red line is a polynomial fit of second order to data points of this work with its $\pm1\sigma$ statistical uncertainty represented by the red shaded area.}
 	\label{fig:DiffCrossPion}
 \end{figure}

 Differential cross sections were determined using the $dp$ elastic luminosities given in Table~\ref{tab:lumresults} and the acceptance-corrected yields for the ${}^3\textrm{He}\,\pi^0$ final state. As an example, the results for the kinetic energy $T=1801 \;\textrm{MeV}$ (i.e.\ supercycle 2, flattop 5) are shown in Fig.~\ref{fig:DiffCrossPion}. The red solid line is a fit of the form
 \begin{equation}
 \label{eq:FitPi0CrossSec}
\frac{d\sigma}{d\Omega} = \sigma_1 + \sigma_2(\cos \vartheta-1) + \sigma_3(\cos \vartheta-1)^2
 \end{equation}
 and the red shaded area describes the $\pm1\sigma$ uncertainty of the fit. The parameters from the fits are given in Table~\ref{tab:Pi0Results} for all beam momenta. Since  the identification of the ${}^3\textrm{He}$ nuclei is exactly the same as for the $\eta$, when using this reaction for the normalization, the systematic uncertainties of the differential cross sections are completely dominated by those of the luminosity given in Table~\ref{tab:lumresults}.

 The only published $dp \rightarrow {}^3\textrm{He}\,\pi^0$ data in our kinematic region were obtained at Saclay in 200~MeV steps at $\cos{\vartheta} = 1$ \cite{Kerboul:1986} and the only one in our energy range was taken at 1800~MeV. This is represented by the blue data point in Fig.~\ref{fig:DiffCrossPion}. In addition to the statistical error, there is an $8\%$ normalization uncertainty in the data of Ref.~\cite{Kerboul:1986} as well as the normalization uncertainty of our data shown in Table~\ref{tab:lumresults}.

\begin{figure}[t]
 	\centering
 	\includegraphics[width=1.\linewidth]{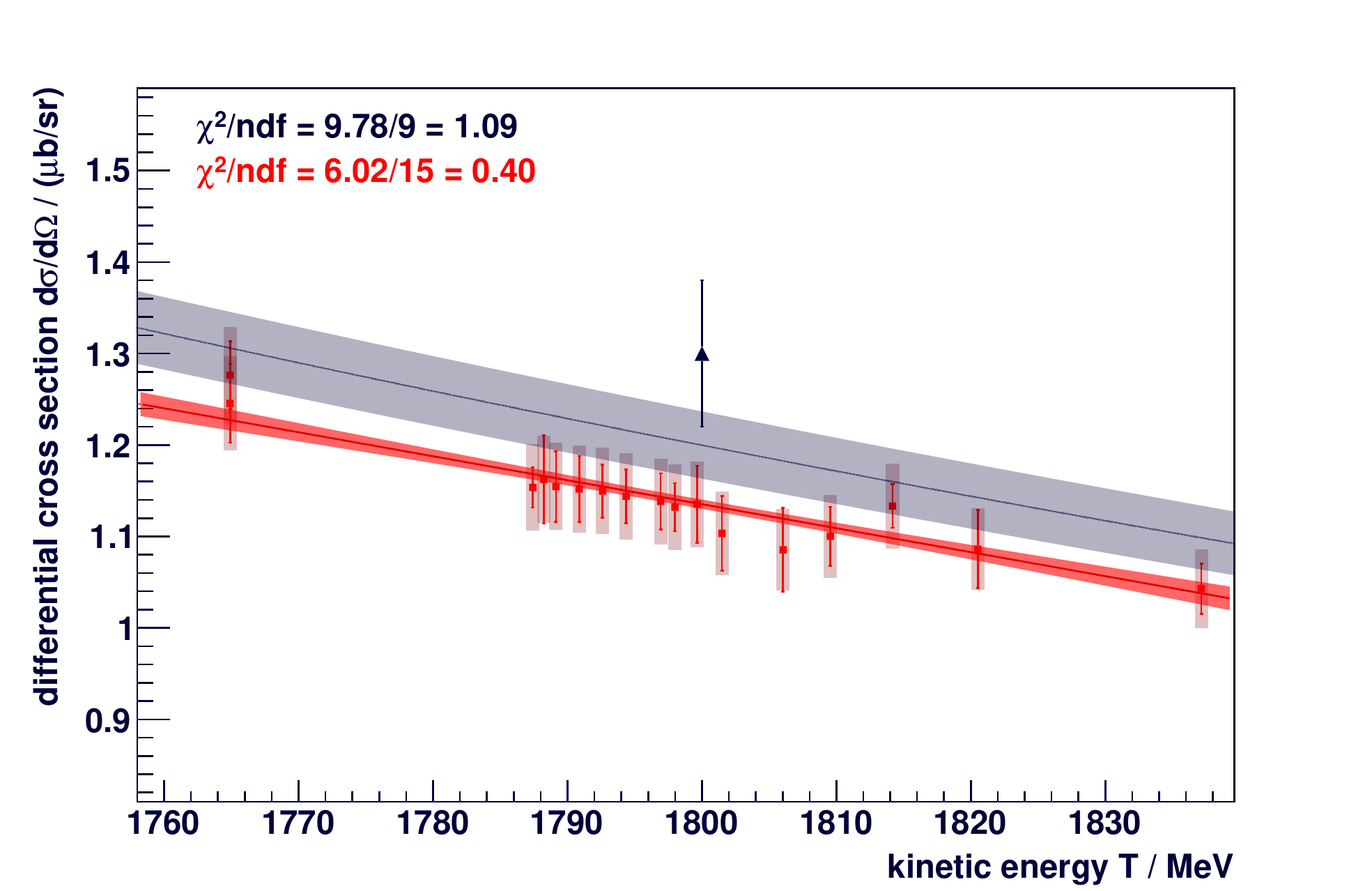}
 	\caption{(Color online) Comparison of the reference data~\cite{Kerboul:1986} (black point) with the results of this work (red points) for the forward $dp \rightarrow {}^3\textrm{He}\,\pi^0$ differential cross section. Here  the red shaded boxes represent the systematic uncertainties. The red line represents a linear fit with its $\pm1\sigma$ statistical uncertainty (red shaded area). A fourth order polynomial fit to the forward reference data~\cite{Kerboul:1986} between $T=1000$~MeV and 2200~MeV is also shown (blue line) with the corresponding $\pm1\sigma$ statistical uncertainty. Since a limited range of $T$ was accessed in this measurement, only one of the Saclay points~\cite{Kerboul:1986} is visible. }
 	\label{fig:KerboulComparision}
 \end{figure}

In order to facilitate a comparison with our results, the Saclay data were fitted with a fourth order polynomial to extract the differential cross sections as a function of $T$, This curve is compared in Fig.~\ref{fig:KerboulComparision} to the extrapolation of our $dp \rightarrow {}^3\textrm{He}\,\pi^0$ results to the forward direction. Given the statistical uncertainty of the interpolation to our $T$ values, the $8\%$ normalization uncertainty of Ref. \cite{Kerboul:1986}, as well as the normalization uncertainty of our data shown in Table~\ref{tab:lumresults}, it is seen that our results are consistent with those obtained at Saclay~\cite{Kerboul:1986}. Just as important, one sees from Fig.~\ref{fig:KerboulComparision} that the slopes of the fit to the Saclay data and that of our results are very similar. This means that any possible systematic energy dependence of the $dp \rightarrow {}^3\textrm{He}\,\eta$ introduced through the use of $dp$ elastic scattering for normalization can be ruled out.

 \begin{table}[t]
 	\centering
 	\caption{Parameters of the second order polynomial fits of Eq.~(\ref{eq:FitPi0CrossSec}) to the $dp \rightarrow {}^3\textrm{He}\,\pi^0$ differential cross sections with their statistical uncertainties.}
 	\label{tab:Pi0Results}
 	\begin{tabular}{c c c c c c c c}
 		\toprule
 		$T$ & $\sigma_1$ & $\Delta\sigma_{1}^{\text{stat}}$ & $\sigma_2$ & $\Delta\sigma_{2}^{\text{stat}}$  & $\sigma_3$ & $\Delta\sigma_{3}^{\text{stat}}$ & $\chi^2/ndf$\\
 		(MeV) &  ($\mu$b\!/\!sr) & ($\mu$b\!/\!sr) & ($\mu$b\!/\!sr) & ($\mu$b\!/\!sr)  &($\mu$b\!/\!sr)  & ($\mu$b\!/\!sr) & \\
 		\specialrule{2pt}{4pt}{6pt}
$1765$	&	1.25	&	0.04	&	10.4	        &	1.2 &	28	&	8	&	2.8	\\
$1765$	&	1.28	&	0.04	&	10.8	        &	1.0	&	31	&	6	&	1.9	\\
$1787$	&	1.15	&	0.02	&	\phantom{1}9.8	&	0.5	&	30	&	4	&	1.7	\\
$1788$	&	1.16	&	0.05	&	\phantom{1}9.8	&	1.4	&	27	&	9	&	3.1	\\
$1789$	&	1.15	&	0.04	&	\phantom{1}9.9	&	1.1	&	29	&	7	&	2.2	\\
$1791$	&	1.15	&	0.04	&	\phantom{1}9.7	&	1.1	&	29	&	7	&	1.9	\\
$1793$	&	1.15	&	0.03	&	10.3	        &	0.8	&	33	&	5	&	1.2	\\
$1794$	&	1.14	&	0.03	&	10.2	        &	0.8	&	32	&	5	&	1.4	\\
$1797$	&	1.14	&	0.03	&	\phantom{1}9.7	&	0.9	&	29	&	5	&	1.5	\\
$1798$	&	1.13	&	0.03	&	\phantom{1}9.4	&	0.7	&	25	&	5	&	1.2	\\
$1800$	&	1.14	&	0.04	&	10.8	        &	1.1	&	38	&	7	&	2.6	\\
$1801$	&	1.10	&	0.04	&	\phantom{1}8.7	&	1.2	&	21	&	8	&	2.9	\\
$1806$	&	1.09	&	0.05	&	\phantom{1}9.1	&	1.4	&	27	&	8	&	2.9	\\
$1810$	&	1.10	&	0.03	&	\phantom{1}8.8	&	0.9	&	23	&	6	&	1.8	\\
$1814$	&	1.13	&	0.02	&	10.7	        &	0.7	&	35	&	4	&	0.8	\\
$1821$	&	1.09	&	0.04	&	\phantom{1}9.9	&	1.2	&	31	&	8	&	2.7	\\
$1837$	&	1.04	&	0.03	&	\phantom{1}9.4	&	0.8	&	30	&	5	&	1.3	\\
 		\bottomrule
 	\end{tabular}
 \end{table}%

%
%
\section{Results}
\label{Results}

The measured luminosities and acceptance-corrected count yields of the ${}^3\textrm{He}\,\eta$ final state were used to determine total and differential cross sections. Table~\ref{tab:TotalCrossSectionEta} shows the resulting total cross sections.
The systematic uncertainties are dominated by those of the normalization based upon deuteron-proton elastic scattering. Neither a variation of the track length limits from $\pm 4\sigma$ to $\pm 3\sigma$ or $\pm 5\sigma$ nor a variation of the energy loss limits influences the results.

Only for the highest flattop at $Q=15.01~\textrm{MeV}$ can one apply also above-threshold data, using the same method to study the background as already described in Sec.~\ref{Analysis}. This is because for data up to $Q=1.63~\textrm{MeV}$ the shifted ${}^3\textrm{He}\,\eta$ signals do not overlap with those seen in the $Q=15.01~\textrm{MeV}$ spectrum. The total cross sections determined using the above-threshold data in the description of the background show a systematic deviation of $\Delta \sigma_{\textrm{sys}} = -5~\textrm{nb}$. This leads to an additional asymmetric uncertainty at 15.01~MeV and this effect has already been included in the uncertainties given in Table~\ref{tab:TotalCrossSectionEta}.
\begin{table}[t]
	\centering
	\caption{Measured total cross sections $\sigma$ of the $dp \to {}^3\textrm{He}\,\eta$ reaction with the statistical uncertainties of the last digits in brackets. The indices $+$ and $-$ on the systematic uncertainties refer to the cases of higher and lower cross sections, respectively.
Also given are the values extracted for the asymmetry parameters $\alpha$ of Eq.~(\ref{alpha}). Though the differential cross sections are generally consistent with a linear dependence on $\cos\vartheta$, at the highest energy the cubic fit of Eq.~(\ref{eq:pol3}) was required and this gave $\beta=-0.008(24)$ and $\gamma=-0.166(51)$. }
	\label{tab:TotalCrossSectionEta}
	\begin{tabular}{c c c c c}
		\toprule
		$Q$ & $\sigma$ & $\Delta\sigma_{\rm{sys}}^{\,+}$ & $\Delta\sigma_{\rm{sys}}^{\,-}$& $\alpha$ \\
		(MeV) & (nb) & (nb) & (nb) &\\
		\specialrule{2pt}{4pt}{6pt}
		$1.14(1)$ &  355(9) & 15 & 15 & 0.048(8)\\
		$1.36(1)$ & 357(9) & 17 & 15 & 0.067(9)\\
		$1.63(1)$ & 357(9) & 16 & 16 & 0.045(8)\\
		$2.10(1)$ & 366(10) & 25 & 15 & \phantom{1}0.068(10)\\
		$2.59(1)$ & 367(10) & 16 & 15 & 0.133(8)\\
		$3.08(1)$ & 371(10) & 17 & 16 & 0.117(8)\\
		$3.79(1)$ & 378(11) & 17 & 15 & \phantom{1}0.107(10)\\
		$4.09(1)$ & 374(10) & 16 & 16 & 0.146(9)\\
		$4.55(1)$ & 379(11) & 16 & 16 & 0.144(9)\\
		$5.07(1)$ & 376(11) & 16 & 16 & 0.179(8)\\
		$6.33(1)$ & 379(11) & 17 & 16 & 0.238(7)\\
		$7.32(1)$ & 388(11) & 15 & 17 & 0.242(9)\\
		$8.60(1)$ & 384(12) & 19 & 16 & 0.310(9)\\
		$10.37(1)$ & 390(12) & 18 & 15 & \phantom{1}0.375(11)\\
		$15.01(1)$ & 403(17) & 18 & 23 & \phantom{1}0.570(34)\\
		\bottomrule
	\end{tabular}
\end{table}%

\begin{figure*}[p]
	\centering
	\includegraphics[width=.92\linewidth, trim = 0cm 1cm 0cm 0cm]{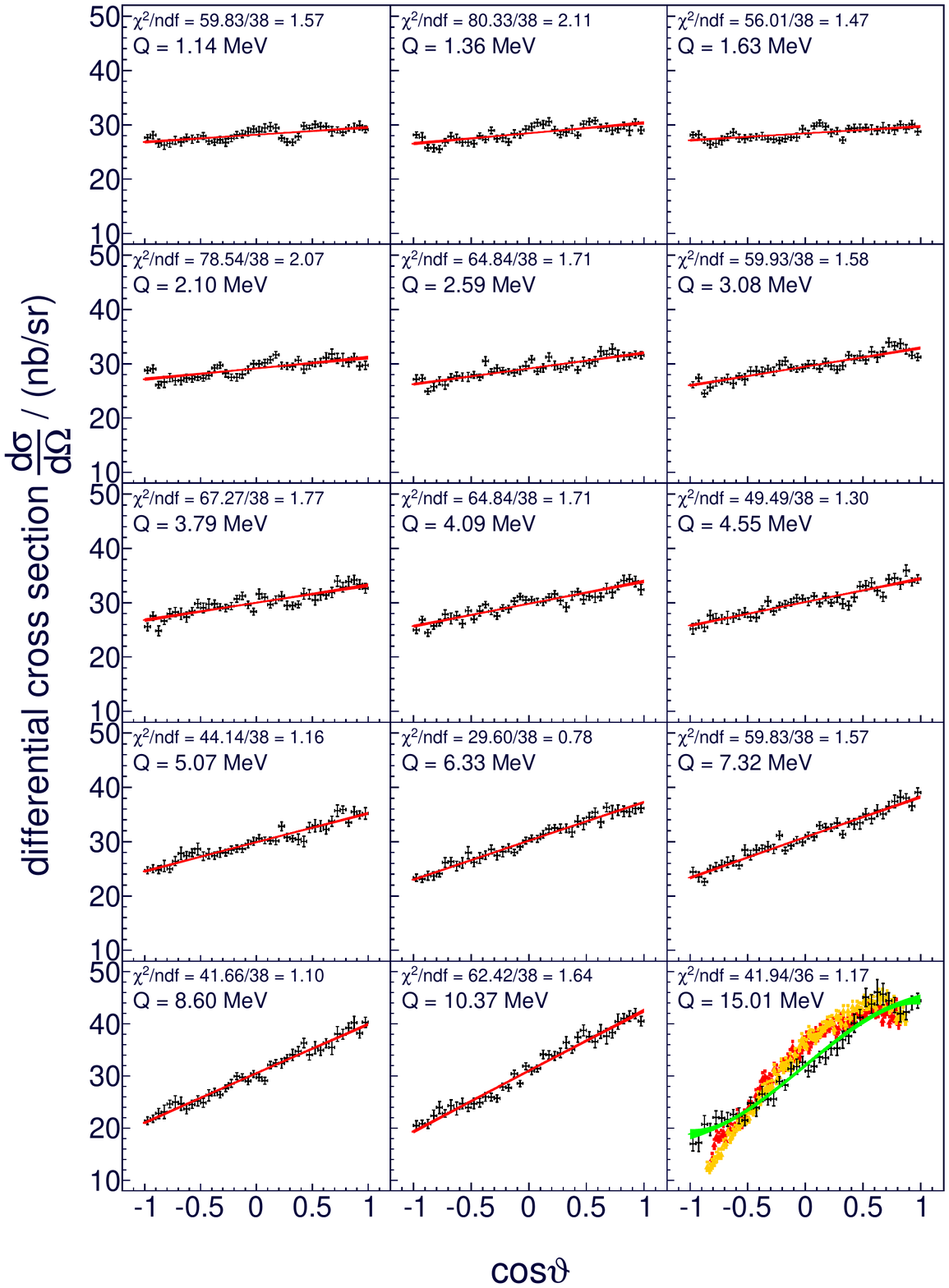}
	\caption{(Color online) Measured differential cross section for all the available beam momentum settings. The red lines are the results of fitting the angular distributions with the linear function of Eq.~\eqref{eq:linear}. The red shaded area represents the corresponding $\pm 1\sigma$ uncertainties. In the case of the highest excess energy, the third degree polynomial of Eq.~\eqref{eq:pol3} (green) was used. The recently published WASA-at-COSY data~\cite{Adlarson:2018}, which were taken at $Q=13.6~\textrm{MeV}$ (yellow) and $Q=18.4~\textrm{MeV}$ (red), are displayed in the final panel. These data have been scaled to the same total cross section as that at 15.01~MeV.}
	\label{fig:DiffCrossSection}
\end{figure*}

Fig.~\ref{fig:DiffCrossSection} shows the differential cross sections $d\sigma/d\Omega$ for all excess energies as well as a linear fit (red line) of the form
\begin{equation}
	\label{eq:linear}
	f(\cos{\vartheta}) = \sigma_0(1 + \alpha \cos{\vartheta})
\end{equation}
and the $\pm 1\sigma$ uncertainty of the fit (red shaded area). In the case of the highest excess energy data point, a third order polynomial fit function of the form
\begin{equation}
	\label{eq:pol3}
		f(\cos{\vartheta}) = \sigma_0 (1  + \alpha \cos{\vartheta} + \beta\cos^2\!{\vartheta}  + \gamma\cos^3\!{\vartheta})
\end{equation}
is displayed (green line), which has a significance of $3.5\sigma$ compared to a linear fit. The results for the fit parameters are given in Table~\ref{tab:TotalCrossSectionEta}. An extensive study of uncertainties was also done here by varying, e.g., the software cut limits. Since the systematic uncertainty of the luminosity, which was the dominant effect for the total cross sections, plays no role for the asymmetry parameter $\alpha$, the systematic uncertainties are roughly only one-fifth of the statistical uncertainties shown in Table~\ref{tab:TotalCrossSectionEta}.

The WASA-at-COSY collaboration recently published data on the $pd\to {}^3\textrm{He}\,\eta$ reaction with a proton beam~\cite{Adlarson:2018}. Their lowest excess energies were $Q=13.6~\textrm{MeV}$ and $Q=18.4~\textrm{MeV}$ and these results, normalized to the total cross section of this work at 15.01~MeV, are compared with our data in the final panel of Fig.~\ref{fig:DiffCrossSection}. Since there are clear differences between the two data sets, a further investigation of the ANKE acceptance was undertaken. In the vicinity of $\vartheta=90^{\circ}$ there was a significant dependence on the azimuthal angle $\phi$, which is not permissable for a two-body reaction involving unpolarized particles. This effect may be due to the $^3$He hitting boundary areas of the ANKE detector where systematic problems with the acceptance corrections are known to occur. In order to analyze this last energy, arbitrary cuts in the $\phi-\vartheta$ plane were made. It should be noted that 15.01~MeV is the only excess energy where there is not full acceptance in the ANKE detector. It is therefore reassuring that for none of the other excess energies does one observe the troubling $\phi$ dependence.

\begin{figure}[!t]
	\centering
	\includegraphics[width=1.\linewidth]{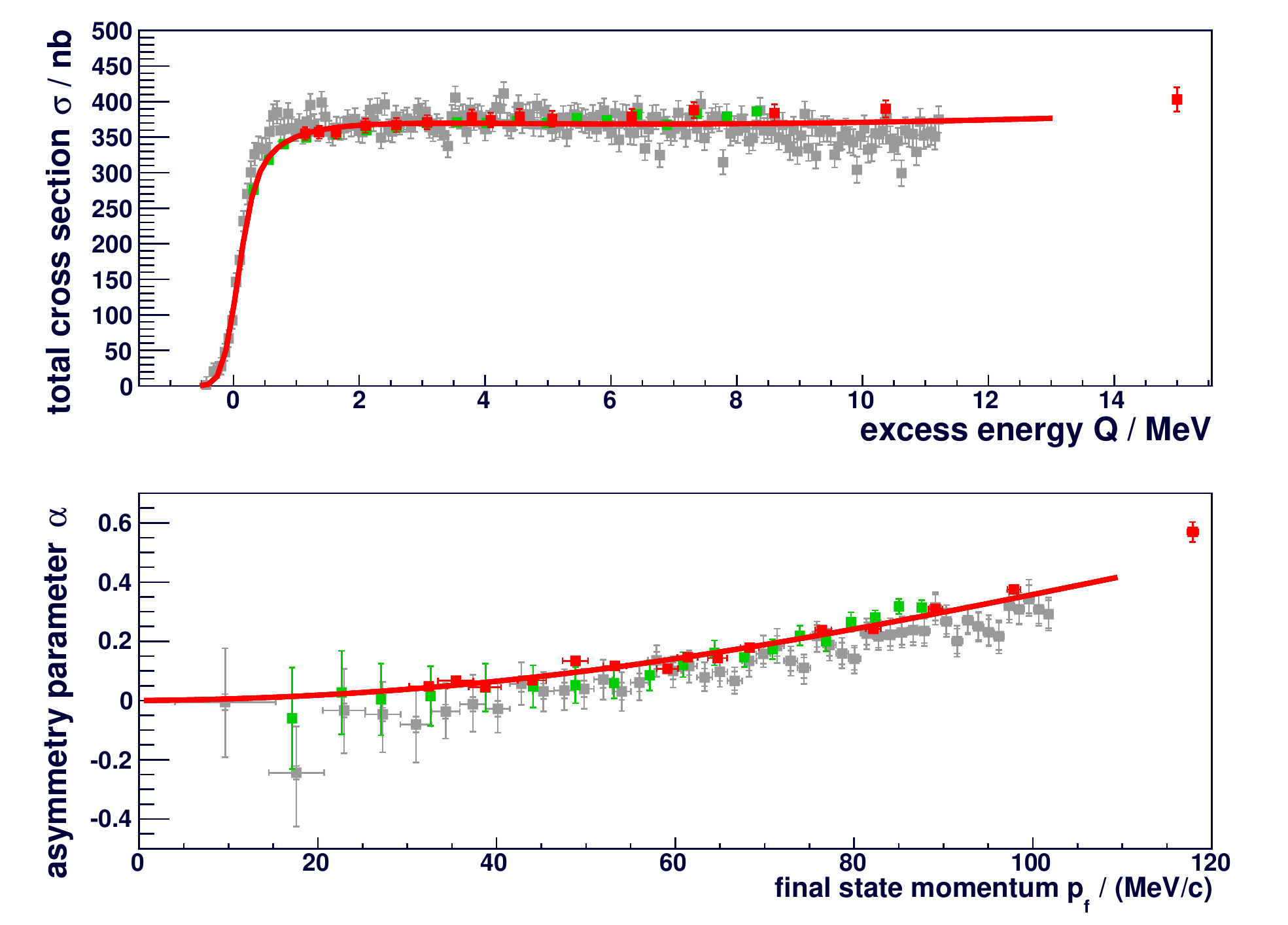}
	\caption{(Color online) Total cross sections and asymmetry parameters determined in a previous ANKE experiment (gray points)~\cite{Mersmann:2007}, COSY-11 measurements (green points)~\cite{Smyrski:2007}, and the results of the current work (red points). Note that the total cross sections of the previous ANKE and COSY-11 data were normalized globally to the data of this work. The red line is a combined fit which fits both spectra simultaneously by using Eqs.~\eqref{eq:totaldiffFit} and \eqref{eq:scatFit}. }
	\label{fig:CombinedFit}
\end{figure}%

Fig.~\ref{fig:CombinedFit} shows the total cross sections as well as the corresponding asymmetry parameters from the previous ANKE (gray) and COSY-11 (green) measurements~\cite{Mersmann:2007,Smyrski:2007}. Given the normalization uncertainties of the different measurements, these earlier total cross sections were renormalized to the present data in order to compare the shapes. The scaling factors were defined by the ratios of the average total cross sections between $Q=1.14~\textrm{MeV}$ and $Q=8.45~\textrm{MeV}$. It can be seen that the new ANKE data (red) also show a plateau-like behavior with a slight tendency to increase with rising beam momentum, which is similar to the COSY-11 data.

\begin{figure}[h]
	\centering
	\includegraphics[width=1.\linewidth]{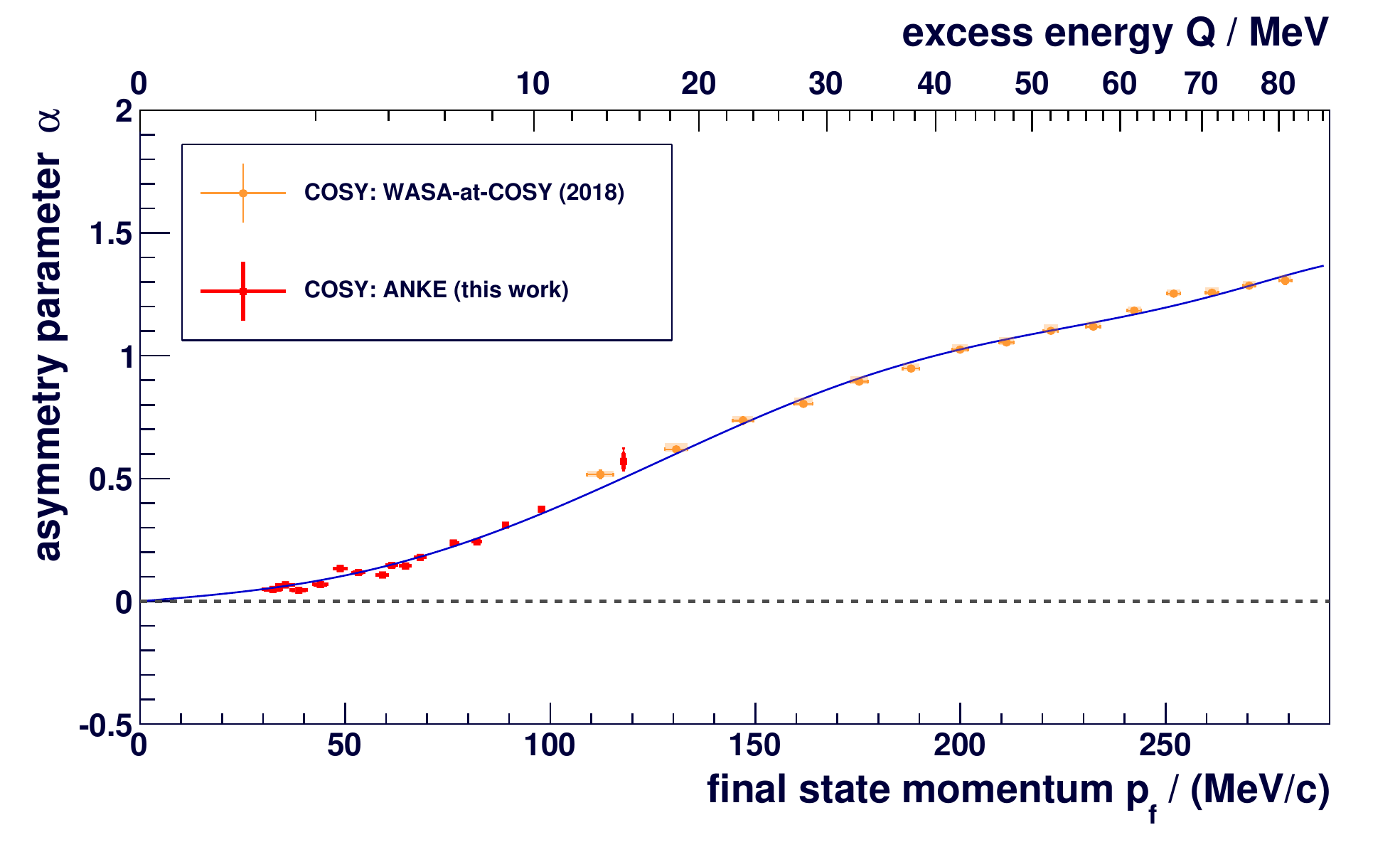}
	\caption{(Color online) Combination of asymmetry parameters determined at two different facilities. The red data points represent the results of this work and the yellow ones results recently published by the WASA-at-COSY collaboration~\cite{Adlarson:2018}. In order to highlight its smooth behavior, the blue line represents an empirical fit to both the ANKE and WASA data sets in odd powers of $p_f$.}
	\label{fig:AngularAsymm}
\end{figure}

There is, of course, no normalization uncertainty in the different measurements of the asymmetry parameters and, given the large error bars of some of the older data, the various results are broadly similar. The high precision of the new ANKE data excludes the possibility that $\alpha$ might change sign for $p_f \lesssim 40~\textrm{MeV}/c$ though all the data do show a monotonically increasing behavior for $p_f \gtrsim 40~\textrm{MeV}/c$. Fig.~\ref{fig:AngularAsymm} shows the values of the asymmetry parameters extracted from this experiment and from that of WASA-at-COSY~\cite{Adlarson:2018}. To emphasize the smooth behavior of this parameter, even in the overlap region of these two experiments, the blue line represents a polynomial fit to the combined data set in terms of odd powers of $p_f$.

%
%
\section{Final state interactions}
\label{FSI}

Since the new ANKE data do not extend to the very near-threshold region of $Q < 1~\textrm{MeV}$, a combined fit was made of the total cross sections and asymmetry parameters measured in the previous ANKE and COSY-11 experiments~\cite{Mersmann:2007,Smyrski:2007} as well as the new ANKE data of this work. For this purpose the total cross sections of the previous ANKE and COSY-11 data were normalized to the new data, as described in the previous section.

If only $s$- and $p$-waves are retained, the observables may be parameterized as
\begin{equation}
	\label{eq:totaldiffFit}
	\sigma = \frac{4\pi p_f}{p_i}\big(  |f_s|^2 + p_f^2|f_p|^2 \big) ~~\textrm{and}~~ \alpha = 2p_f\frac{\Re(f_s^*f_p)}{|f_s|^2+p_f^2|f_p|^2}
\end{equation}
with the effective $p$-wave amplitude $f_p$ being taken as a complex constant.

In the previous ANKE analysis~\cite{Mersmann:2007}, the $s$-wave amplitude was taken as the product of two poles, it being argued that the second pole was so far away that this would have little influence on the parameters deduced for the near-threshold pole. The analysis is on much firmer theoretical grounds if one describes the final state interaction in terms of a Jost function that goes to a non-zero constant for large $p_f$~\cite{Newton:1982}. Thus we take
\begin{equation}
	\label{eq:scatFit}
	f_s = f_s^0 \left(\frac{ 1-p_f/p_2}{1-p_f/p_1}\right).
\end{equation}

A combined fit was made to the total cross section and asymmetry parameter using the functions of Eqs.~\eqref{eq:totaldiffFit} and \eqref{eq:scatFit}. A fit range of $p_f \leq 110~\textrm{MeV}/c$ was chosen to minimize contributions from higher partial waves. It should also be noticed that, in addition, as $p_f$ increases the $s$-wave final state interaction becomes less important. This fit range leads to the parameters
\begin{equation}
	\begin{split}
		f_s^0 &= 49(1)~\textrm{(nb/sr)}^{1/2} \\
		p_1 &= (-33(5) \pm i16(5))~\textrm{MeV}/c \\
		p_2 &= (-56(46) \mp i225(23))~\textrm{MeV}/c \\
		\chi^2/\textit{ndf} &= 497.40/307 = 1.62,
	\end{split}
\end{equation}
where, it should be stressed, the uncertainties are strongly correlated. This is to be compared to the two-pole description of the earlier ANKE data given in Ref.~\cite{Wilkin:2007}, which had $p_1=(-4(7) \pm i19(2))~\textrm{MeV}/c$. The real parts of $p_1$ differ by large amounts compared to the quoted errors but these do not include any uncertainty arising from the FSI assumption in Eq.~\eqref{eq:scatFit} and the corresponding one in Ref.~\cite{Mersmann:2007}. This difference in $p_1$ leads to one in the pole position, viz., $Q=(0.91(0.40)\mp i1.15(0.40))$~MeV versus $Q=(-0.35(0.13)\pm i0.21(0.29))$~MeV~\cite{Mersmann:2007}. The changes in the value of $Q$ are significant in the question of the possible existence of a $_{\eta}^{3}$He mesic nucleus.

The form used in Eq.~\eqref{eq:scatFit} corresponds completely to that of a complex Bargman potential~\cite{Bargmann:1949} for which the $\eta\,^{3}$He scattering length becomes
\begin{equation}
	a = i(p_2-p_1)/(p_1p_2). \quad
\end{equation}
 The value resulting from our measurements is $a = (\pm(3.2\pm0.8) - i(4.7\pm2.0))~\textrm{fm}$. 

%
%
\section{Summary}
\label{Summary}

The differential cross section for the $dp\to {}^3\textrm{He}\,\eta$ reaction was measured with the ANKE magnetic spectrometer at 15 different excess energies from $Q=1.14~\textrm{MeV}$ to $15.01~\textrm{MeV}$. The $^3$He was clearly identified in the spectrometer and the meson determined from the missing mass in the reaction. After making acceptance corrections, the normalization (i.e., the luminosity $L$) was determined from deuteron-proton elastic scattering that was measured in parallel. This led to a systematic precision of $\Delta L_{\rm sys} \approx 6\%$ and statistical precision $\Delta L_{\rm stat} \approx 1\%$, which are improvements by at least a factor of two compared to the earlier measurements at ANKE~\cite{Mersmann:2007,Rausmann:2009}. The normalization could be checked through the measurement of the $dp\to {}^3\textrm{He}\,\pi^0$ differential cross section at small angles, though this method is hindered by the 8\% systematic uncertainty in the only available reference data~\cite{Kerboul:1986}. Nevertheless, these results do show that the energy dependence of the normalization achieved through $dp$ elastic scattering is broadly correct. As a byproduct of these $\eta$ production data, we have obtained well normalized $dp\to {}^3\textrm{He}\,\pi^0$ differential cross section in small energy steps around a deuteron beam energy of $T=1800$~MeV.

Due to full geometrical coverage over the entire $\cos\vartheta$ range and the high event rates, angular distributions with a bin width of $\Delta\cos\vartheta = 0.05$ could be extracted for each beam momentum setting. Apart from the highest excess energy, the differential cross sections seem to be linear in $\cos\vartheta$, with a $90^{\circ}$ slope that increases with rising $Q$. However, at 15.01~MeV a third order polynomial was needed to describe the data and this is a sign that higher partial waves are present.

The asymmetry parameters $\alpha$ vary smoothly from $Q\leq 15.01$~MeV in our measurements at ANKE to the $Q\geq 13.6$~MeV at WASA~\cite{Adlarson:2018}. Taken together, the data show $\alpha$ in unprecedented detail up to an excess energy of $Q=80.9~\textrm{MeV}$. Furthermore, the possible change of sign of the asymmetry parameter $\alpha$ near threshold, indicated in earlier measurements, is not confirmed by the new ANKE data. Fitting the low energy total cross section and asymmetry parameter with an $s$-wave FSI function that has a more believable large $p_f$ behaviour changes the position of the near-threshold pole and this is important in the discussions of $\eta$-mesic nuclei, of which the case of $_{\eta}^3$He studied here is probably the best example. It is reassuring that the value of the scattering length estimated using an optical potential ansatz is $a = (2.2\pm1.3 - i(4.9\pm0.6))$~fm\cite{Xie:2017}, which is compatible with our experimental result of $a = (\pm(3.2\pm0.8) - i(4.7\pm2.0))$~fm. Other reactions are less favored and, to illustrate this point, the only similar possible signal is found in the $dd\to {}^4\textrm{He}\,\eta$ reaction~\cite{Willis:1997} and the cross section is about a factor of 50 lower than that measured here. Other above-threshold experiments involving heavier nuclei give even smaller values.

Above-threshold measurements cannot distinguish between bound and antibound (virtual) states but searches below threshold for signs of an $_{\eta}^3$He, such as that carried out at WASA~\cite{Adlarson:2017}, are hampered by the necessary absence of an $\eta$ signal. In 1988 no convincing sign was found for $\eta$-mesic nuclei in below-threshold measurements~\cite{Chrien:1988} and this is still the situation now.

We turn finally to attempts to provide a theoretical description of the $dp\to {}^3\textrm{He}\,\eta$ reaction. Due to the large mass of the $\eta$ meson, a single-nucleon mechanism is much reduced in importance and it is likely that both nucleons in the deuteron must be involved dynamically~\cite{Laget:1988}. A simple semi-classical model based on these ideas was proposed~\cite{Kilian:1990} and later put in quantum mechanical form~\cite{Kondratyuk:1995,FaeldtWilkin:1995}. Although this approach has some success at threshold, it fails to describe the energy dependence and the shapes of the cross section away from threshold. There is therefore much theoretical work to do, especially to connect the reaction mechanism with the $\eta\,^3\textrm{He}$ final state interaction.

%
%
\section*{Acknowledgments}

The authors express their thanks to the COSY machine crew for producing such good experimental conditions and also to the other members of the ANKE collaboration for diverse help in the experiment. This work was supported in part by the JCHP Fremde Forschungs und Entwicklungsarbeiten (FFE) of the Forschungszentrum J\"{u}lich and also by the DFG through the Research Training Group GRK2149.
%
%


\begin{thebibliography}{99}
%
\bibitem{Berger:1988} J.~Berger \emph{et al.}, Phys.\ Rev.\ Lett.\ \textbf{61} 919 (1988).
%
\bibitem{Wilkin:1993} C.~Wilkin, Phys.\ Rev.\ C \textbf{47}, R938 (1993).
%
\bibitem{Bhalerao:1985cr} R.S.~Bhalerao and L.C.~Liu, Phys.\ Rev.\ Lett.\ \textbf{54}, 865 (1985).
%
\bibitem{Haider:1986EtaMesicNuclei} Q.~Haider and L.C.~Liu, Phys.\ Lett.\ B \textbf{172}, 257 (1986).
%
\bibitem{Pfeiffer:2004} M.~Pfeiffer \emph{et al.}, Phys.\ Rev.\ Lett.\ \textbf{92}, 252001 (2004).
%
\bibitem{Pfeiffer:2005} M.~Pfeiffer \emph{et al.}, Phys.\ Rev.\ Lett.\ \textbf{94}, 049102 (2005).
%
\bibitem{Pheron:2012} F.~Pheron \emph{et al.}, Phys.\ Lett.\ B \textbf{709}, 21 (2012).
%
\bibitem{Mersmann:2007} T.~Mersmann \emph{et al.}, Phys.\ Rev.\ Lett.\ \textbf{98}, 242301 (2007).
%
\bibitem{Smyrski:2007} J.~Smyrski \emph{et al.}, Phys.\ Lett.\ B \textbf{649}, 258 (2007).
%
\bibitem{Adlarson:2018} P.~Adlarson \emph{et al.}, Phys.\ Lett.\ B \textbf{782}, 297 (2018).
%
\bibitem{Mayer:1995} B.~Mayer \emph{et al.}, Phys.\ Rev.\ C \textbf{53}, 2068 (1996).
%
\bibitem{Adam:2007} H.-H.~Adam \emph{et al.}, Phys.\ Rev.\ C \textbf{75}, 014004 (2007).
%
\bibitem{Wilkin:2007} C.~Wilkin \emph{et al.}, Phys.\ Lett.\ B \textbf{654}, 92 (2007).
%
\bibitem{Goslawski:2010} P.~Goslawski \emph{et al.}, Phys.\ Rev.\ ST Accel.\ Beams \textbf{13}, 022803 (2010).
%
\bibitem{Goslawski:2012} P.~Goslawski \emph{et al.}, Phys.\ Rev.\ D \textbf{85}, 112011 (2012).
%
\bibitem{Rausmann:2009} T.~Rausmann \emph{et al.}, Phys.\ Rev.\ C \textbf{80}, 017001 (2009).
%
\bibitem{Fritzsch:2018} C.~Fritzsch \emph{et al.}, Phys.\ Lett.\ B \textbf{784}, 277 (2018).
%
\bibitem{Fritzsch:2019} C.~Fritzsch, PhD thesis, Westf{\"a}lische Wilhelms-Universit{\"a}t, M{\"u}nster (2019).
%
\bibitem{Barsov:2001} S.~Barsov \emph{et al.}, Nucl.\ Instrum.\ Meth.\ Phys.\ Res.\ A \textbf{462}, 364 (2001).
%
\bibitem{Khoukaz:1999} A.~Khoukaz \emph{et al.}, Eur.\ Phys.\ J.\ D \textbf{5}, 275 (1999).
%
\bibitem{Tanabashi:1898} M.~Tanabashi \emph{et al.}, Phys.\ Rev.\ D \textbf{98}, 030001 (2018).
%
\bibitem{Dalkhazhav:1969} N.~Dalkhazhav \emph{et al.}, Sov.\ J.\ Nucl.\ Phys.\ \textbf{8}, 196 (1969).
%
\bibitem{Winkelmann:1980} E.~Winkelmann \emph{et al.}, Phys.\ Rev.\ C \textbf{21}, 2535 (1980).
%
\bibitem{Irom:1984} F.~Irom \emph{et al.}, Phys.\ Rev.\ C \textbf{28}, 2380 (1983).
%
\bibitem{Velichko:1988} G.N.~Velichko \emph{et al.}, Sov.\ J.\ Nucl.\ Phys.\ \textbf{47}, 755 (1988).
%
\bibitem{Guelmez:1991} E.~Guelmez \emph{et al.}, Phys.\ Rev.\ C \textbf{43}, 2067 (1991).
%
\bibitem{Kerboul:1986} C.~Kerboul \emph{et al.}, Phys.\ Lett.\ B \textbf{181}, 28 (1991).
%
\bibitem{Newton:1982} R.~G.~Newton, \emph{Scattering Theory of Waves and Particles}, (Springer-Verlag, New York, 1982).
%
\bibitem{Bargmann:1949} V.~Bargmann, Phys.\ Rev.\ \textbf{75}, 301 (1949).
%
\bibitem{Xie:2017} Ju-Jun Xie \emph{et al.}, Phys.\ Rev.\ C \textbf{95}, 015202 (2017).
%
\bibitem{Willis:1997} N.~Willis \emph{et al.}, Phys.\ Lett.\ B \textbf{406}, 14 (1997).
%
\bibitem{Adlarson:2017} P.~Adlarson \emph{et al.}, Nucl.\ Phys.\ A \textbf{959}, 102 (2017).
%
\bibitem{Chrien:1988} R.E.~Chrien \emph{et al.}, Phys.\ Rev.\ Lett.\ \textbf{60}, 2595 (1988).
%
\bibitem{Laget:1988} J.M.~Laget and J.F.~Lecolley, Phys.\ Rev.\ Lett.\ \textbf{61} 2069 (1988).
%
\bibitem{Kilian:1990} K.~Kilian and H.~Nann, AIP Conf.\ Proc.\ \textbf{221}, 185 (1991).
%
\bibitem{Kondratyuk:1995} L.A.~Kondratyuk, A.V.~Lado, and Yu.N.~Uzikov, Phys.\ At Nucl.\ \textbf{58}, 473 (1995).
%
\bibitem{FaeldtWilkin:1995} G.~F\"aldt and C.~Wilkin, Nucl.\ Phys.\ A \textbf{587}, 769 (1995).
%
\end{thebibliography}
\end{document}